\documentclass[pra, aps, reprint, graphics, floatfix, tightenlines, twocolumn, longbibliography, superscriptaddress]{revtex4-1}

\pdfoutput=1

\usepackage{epsfig}
\usepackage{amsmath,amssymb,amsfonts}
\usepackage{mathrsfs}
\usepackage{bbm}
\usepackage{slashed}
\usepackage{graphicx}
\usepackage{verbatim}
\usepackage[usenames]{color}
\usepackage{mathtools}
\usepackage[ruled,linesnumbered,commentsnumbered]{algorithm2e}
\usepackage{float}
\usepackage[caption=false]{subfig}

\usepackage{appendix}
\usepackage{booktabs}
\usepackage{relsize}
\usepackage{algpseudocode}
\usepackage[siunitx]{circuitikz}
\usepackage[section]{placeins}
\usepackage{tikz}
\usetikzlibrary{shadings}

\usepackage{lipsum} % remove later, used for placeholder texts

\usepackage{enumerate}

\usepackage{url}
\usepackage{hyperref}
\hypersetup{
    colorlinks=true,
    linkcolor=blue,
    citecolor=blue, 
    filecolor=magenta,      
    urlcolor=blue,
}

\usepackage{units}

\makeatletter
\def\@bibdataout@aps{%
\immediate\write\@bibdataout{%
@CONTROL{%
apsrev41Control%
\longbibliography@sw{%
    ,author="08",editor="1",pages="1",title="0",year="1"%
    }{%
    ,author="08",editor="1",pages="1",title="",year="1"%
    }%
  }%
}%
\if@filesw \immediate \write \@auxout {\string \citation {apsrev41Control}}\fi 
}
\makeatother

\usepackage{mleftright} %to get rid of the extra space around parentheses with \left and \right: write \mleft, \mright instead

\newcommand{\ket}[1]{\mleft|#1\mright\rangle}

\newcommand{\figref}[1]{\mbox{Fig.~\ref{#1}}}
\newcommand{\tabref}[1]{\mbox{Table~\ref{#1}}}
\newcommand{\secref}[1]{\mbox{Section~\ref{#1}}}

\renewcommand{\eqref}[1]{\mbox{Eq.~(\ref{#1})}}

\newcommand{\figpanel}[2]{Fig.~\hyperref[#1]{\ref*{#1}(#2)}}
\newcommand{\figpanels}[3]{Fig.~\hyperref[#1]{\ref*{#1}(#2)-(#3)}}
\newcommand{\figpanelNoPrefix}[2]{\hyperref[#1]{\ref*{#1}(#2)}}
\usepackage{mleftright} %to get rid of the extra space around parentheses with \left and \right: write \mleft, \mright instead 

\definecolor{myblue}{RGB}{71,127,240}
\definecolor{myred}{RGB}{255,45,72}

\usepackage[nolist]{acronym}

\begin{document}
%%%%%%%%%%%%%%%%%%%%%%%%%%%%%%%%%%%%%%%%%%%%%%%%%%%%%%%%%%%%%%%%%%%%%%%%%%%%%%%%%%%%
% Acronyms
%%%%%%%%%%%%%%%%%%%%%%%%%%%%%%%%%%%%%%%%%%%%%%%%%%%%%%%%%%%%%%%%%%%%%%%%%%%%%%%%%%%%

\captionsetup[subfigure]{labelformat=empty}

\begin{acronym}
    \acro{MWPM}{Minimum Weight Perfect Matching}
\end{acronym}

\newacro{MWPM}{Minimum Weight Perfect Matching}
\newacro{CNN}{convolutional neural network}
\newacro{ReLU}{rectified linear unit}
\newacro{SGD}{stochastic gradient descent}
\newacro{RL}{reinforcement learning}
\newacro{DRL}{deep reinforcement learning}
\newacro{NN}{neural network}
\newacro{ML}{machine learning}

%%%%%%%%%%%%%%%%%%%%%%%%%%%%%%%%%%%%%%%%%%%%%%%%%%%%%%%%%%%%%%%%%%%%%%%%%%%%%%%%%%%%
% Title page 
%%%%%%%%%%%%%%%%%%%%%%%%%%%%%%%%%%%%%%%%%%%%%%%%%%%%%%%%%%%%%%%%%%%%%%%%%%%%%%%%%%%%

\title{Error-rate-agnostic decoding of topological stabilizer codes} 

\author{Karl Hammar}
\affiliation{Department of Physics, University of Gothenburg, SE-41296 Gothenburg, Sweden}

\author{Alexei Orekhov}

\affiliation{Department of Microtechnology and Nanoscience, Chalmers University of Technology, SE-41296 Gothenburg, Sweden}

\author{Patrik Wallin Hybelius}

\affiliation{Department of Physics, University of Gothenburg, SE-41296 Gothenburg, Sweden}

\author{Anna Katariina Wisakanto}

\affiliation{Department of Physics, University of Gothenburg, SE-41296 Gothenburg, Sweden}

\author{Basudha Srivastava}
\affiliation{Department of Physics, University of Gothenburg, SE-41296 Gothenburg, Sweden}

\author{Anton Frisk Kockum}
\affiliation{Department of Microtechnology and Nanoscience, Chalmers University of Technology, SE-41296 Gothenburg, Sweden}

\author{Mats Granath}
\email[]{mats.granath@physics.gu.se}
\affiliation{Department of Physics, University of Gothenburg, SE-41296 Gothenburg, Sweden}

\begin{abstract}

Efficient high-performance decoding of topological stabilizer codes has the potential to crucially improve the balance between logical failure rates and the number and individual error rates of the constituent qubits. High-threshold maximum-likelihood decoders require an explicit error model for Pauli errors to decode a specific syndrome, whereas lower-threshold heuristic approaches such as minimum weight matching are ``error agnostic''. Here we consider an intermediate approach, formulating a decoder that depends on the  bias, i.e., the relative probability of phase-flip to bit-flip errors, but is agnostic to error rate. Our decoder is based on counting the number and effective weight of the most likely error chains in each equivalence class of a given syndrome. We use Metropolis-based Monte Carlo sampling to explore the space of error chains and find unique chains, that are efficiently identified using a hash table. Using the error-rate invariance the decoder can sample chains effectively at an error rate which is higher than the physical error rate and without the need for ``thermalization'' between chains in different equivalence classes. 
Applied to the surface code and the XZZX code, the decoder matches maximum-likelihood decoders for moderate code sizes or low error rates. We anticipate that, because of the compressed information content per syndrome, it can be taken full advantage of in combination with machine-learning methods to extrapolate Monte Carlo-generated data.

\end{abstract}	
\maketitle

%%%%%%%%%%%%%%%%%%%%%%%%%%%%%%%%%%%%%%%%%%%%%%%%%%%%%%%%%%%%%%%

\section{Introduction}

Quantum decoherence is one of the major challenges that has to be overcome in building a quantum computer~\cite{PhysRevA.52.R2493, PhysRevLett.77.793, gottesman1997stabilizer,Nielsen2000, terhal2015quantum, Girvin2021}. A prominent line of research to address this issue focuses on topological stabilizer codes implemented on low-connectivity lattices of qubits~\cite{kitaev2003fault, bravyi1998quantum, dennis2002topological, Raussendorf2007, fowler2012surface}, with small stabilizer codes presently being realized experimentally~\cite{Kelly2015, Takita2017, CKAndersen2020, marques2022logical, Chen2021Google, Erhard2021, satzinger2021realizing, Egan2021, Ryan-Anderson2021, Livingston2021, postler2021demonstration, Krinner2021, Bluvstein2021}.
% \cite{Gong2021}
The topological codes provide protection against errors by encoding the quantum information into entangled states that are removed from each other by a code distance $d$, corresponding to the minimal number of local qubit operations required for a logical operation. These logical operators act as effective Pauli operators on the logical code space. Local stabilizers in the form of local Pauli operations commute with the logical operators; measuring these stabilizers provides a syndrome of the possible error configurations, so-called error chains, that are affecting the code. The error chains can be grouped into equivalence classes depending on whether they commute with a given representation of the logical operators or not.   

To decide which corrective actions to take given a measured syndrome, there is a range of different decoders. A maximum-likelihood decoder (MLD) aims directly at the core problem of decoding a stabilizer code, which is to identify the most likely equivalence class of error chains that correspond to the syndrome, thus specifying a correction chain (or class of chains) least likely to cause a logical error~\cite{dennis2002topological}. Ideally, an MLD will be optimal in the sense of providing the highest possible logical fidelity given the type of stabilizer code and the error rate of individual qubits. Consequently, it will also provide the highest possible threshold for the code~\cite{dennis2002topological, wang2003confinement, PhysRevX.2.021004, katzgraber2013stability}, corresponding to the error rate below which the logical failure rate can always be decreased by increasing the code size.

In contrast to maximum-likelihood decoding, there are ``heuristic'' decoders that suggest the correction chain based on a rule or algorithm which is not guaranteed to fall in the most likely equivalence class. The standard such algorithm is minimum weight perfect matching (MWPM)~\cite{edmonds1965paths, dennis2002topological, wang2009threshold, PhysRevA.83.020302, PhysRevLett.102.200501, PhysRevA.81.022317, fowler2013optimal, delfosse2014decoding, fowler2015minimum, criger2018multi}, which is based on finding a most likely error chain by mapping the syndrome to a graph problem with weighted edges. Other examples are based on addressing the syndrome successively from small to large scale, using the renormalization group \cite{duclos2010fast}, cellular-automaton \cite{herold2015cellular,kubica2018cellular}, or the union-find algorithm \cite{delfosse2017almost,PhysRevA.102.012419}. Such heuristic decoders generally have higher logical failure rates and lower error thresholds than MLDs. Nevertheless, because of the computational complexity of MLDs, non-optimal decoders may be preferable, or even necessary, for practical purposes~\cite{fowler2012surface}. 

Following the recent transformative developments in deep learning~\cite{lecun2015deep, goodfellow2016deep}, with applications to quantum physics~\cite{carleo2017solving, carrasquilla2017machine, van2017learning, carrasquilla2020machine, Ahmed2021}, decoders based on machine-learning algorithms have also been developed. Approaches using deep reinforcement learning~\cite{sweke2020reinforcement, andreasson2018quantum, nautrup2018optimizing, colomer2020reinforcement, Fitzek_DRL} or evolutionary algorithms~\cite{10.21468/SciPostPhys.11.1.005} have the appealing property that they can learn by exploring a model of the errors and measurements of the physical code, but a disadvantage is that they are difficult to train and to scale to large code distances. Deep-learning approaches based on supervised learning~\cite{torlai2017neural, krastanov2017deep, varsamopoulos2017decoding, baireuther2018machine, breuckmann2018scalable, chamberland2018deep, Ni2020neuralnetwork, gicev2021scalable} have the potential for fast and scalable decoding. However, supervised learning is based on training data, which has to be generated using a reference decoder, and the type and quality of the data will be crucial. Part of the objective of the present work is to generate data in a form which is compact enough for deep learning while still containing information that allows for high performance and flexibility with respect to error rate. 
    %\cite{theveniaut2021neat}

Two types of MLDs for general noise models are the Metropolis-based Markov-chain Monte Carlo (MCMC) algorithm by Wootton {\em et al.}~\cite{wootton2012, hutter2014efficient} and the matrix-product-states (MPS) decoder by Bravyi {\em et al.}~\cite{Bravyi2014}. %Both of these methods are approximate in practice from limiting the length of Markov chains or the bond dimension of the MPS, but give error thresholds that are very close to the theoretical limits. 
Apart from the high time complexity, a limitation of these decoders is that they are fixed to the noise model and error rate, i.e., the specific single-qubit error rates $p_x$, $p_y$, and $p_z$ for $X$, $Y$ and $Z$ Pauli errors, with $p=p_x+p_y+p_z$ the total error rate. Given a syndrome $s$ consistent with a set of chains $\{C\}_s$, the probability $P_E$ that this is caused by an error chain in equivalence class $E$ is proportional to the sum of the (a priori) probabilities of the chains, $\pi_C$, i.e., $P_E\sim\sum_{C\in E}\pi_C$. The MCMC decoder is based on Metropolis sampling of the error chains, while the MPS decoder performs an efficient truncated summation of the probabilities. For both of these methods, a statement about most likely equivalence class for a given syndrome does not generalize beyond the error rates for which the calculation was performed. 

In this article, we explore a decoder based on a highly reduced count of the error chains. We only consider the most likely set of chains and their degeneracy in each equivalence class, in order to approximate the probability of the equivalence class. Defining an effective weight $w$, which takes into account the relative probability of $X$, $Y$, and $Z$ errors, the set of most likely chains in each class is independent of the overall error rate. The decoder is in this sense ``error-rate agnostic''. For codes with a single logical qubit and four corresponding equivalence classes of error chains, each syndrome is characterized by an octet of numbers corresponding to weight and degeneracy of the most likely (`lightest') chains in each class. In contrast, an ideal MWPM decoder would identify one lightest chain among all the equivalence classes, whereas the MCMC decoder would take into account a large number of error chains with varying weights. We find that for moderate code distances, the limited information of weight and degeneracy is sufficient to provide near-optimal decoding, matching results from the MCMC MLD for different noise bias for two standard topological stabilizer codes.  

We anticipate two potential uses for this ``effective weight and degeneracy'' (EWD) decoder. The first is as a stand-alone near-optimal decoder. While using Metropolis sampling to find high probability chains, the decoder has comparatively faster convergence than the MCMC decoder as equivalence classes can be explored separately and at a higher error rate than the physical error rate. The second use is to provide high information-content data for supervised deep-learning-based decoders that may allow for flexibility with respect to the overall error rate.  

This article is organized as follows. We start in \secref{background} with a short review of the two stabilizer codes that we will use to benchmark our decoder. In \secref{model}, we describe the basic formulation of the decoder. After this, in \secref{results}, we show some test cases and compare to the MCMC decoder, the MPS decoder, as well as exact analytical results. This is followed by a discussion in \secref{discussion} and a conclusion in \secref{conclusion}.

%%%%%%%%%%%%%%%%%%%%%%%%%%%%%%%%%%%%%%%%%%%%%%%%%%%%%%%%%%%%%%%

\section{Topological stabilizer codes}
\label{background}

To investigate the performance of the EWD decoder, we consider two different stabilizer codes: the rotated surface code and the XZZX code. The surface code~\cite{kitaev2003fault, bravyi1998quantum, dennis2002topological} is of the Calderbank-Shor-Steane (CSS) type with stabilizers containing only one type of Pauli operators~\cite{Nielsen2000, terhal2015quantum}. (Formally, these are the generators of the stabilizer group. Any product of the generators is also a stabilizer.) The rotated configuration reduces the number of qubits without reducing the code distance~\cite{PhysRevA.76.012305, PhysRevX.9.041031}. Recently it was pointed out~\cite{bonilla2020xzzx} that a modified version of the code~\cite{PhysRevLett.90.016803,PhysRevLett.107.270502}, with mixed `XZZX' stabilizers, has very high threshold for biased noise, and also allows for efficient decoding. For depolarizing noise the two codes are equivalent, but for $Z$- or $X$-biased noise they have very different properties; whereas the threshold decreases with bias for the surface code it increases for the XZZX code. 

\begin{figure}
    \begin{tikzpicture}
    \node[circle, draw=white, draw opacity=0, text=black, scale=1.2] (a) at (-3.7, 3.7) {(a)};
    \node[inner sep=0pt] (y) at (0,0)
    {\includegraphics[width=0.43\textwidth]{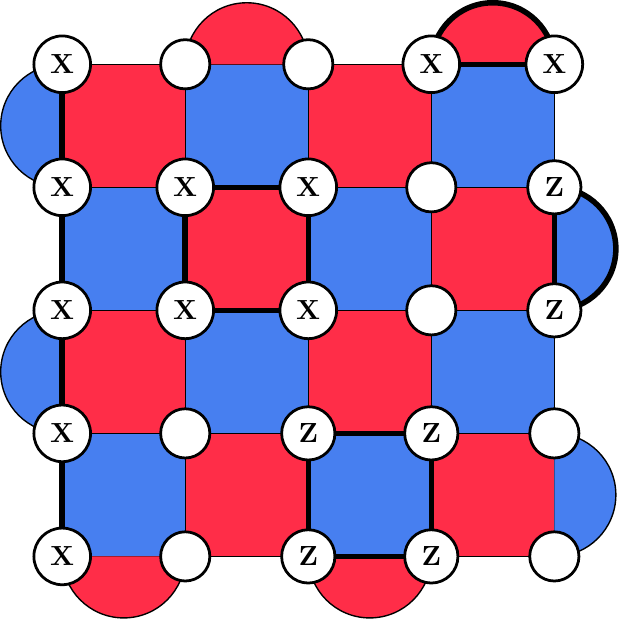}};
    \end{tikzpicture}
    \begin{tikzpicture}
    \node[circle, draw=white, draw opacity=0, text=black, scale=1.2] (b) at (-3.7, 3.7) {(b)};
    \node[inner sep=0pt] (y) at (0,0)
    {\includegraphics[width=0.43\textwidth]{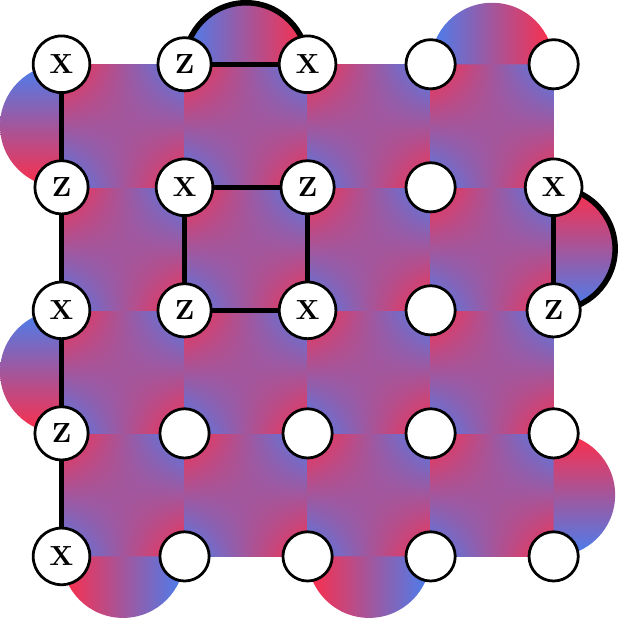}};
    \end{tikzpicture}
    \caption{Illustrations of the two stabilizer codes used to benchmark our EWD decoder.
    (a) The rotated surface code for $d=5$, showing weight-4 and weight-2 stabilizers acting on a grid of $d^2$ physical qubits (circles). The logical operators used to define the equivalence classes of error chains are weight-5 strings: $X_L$, on the left edge, and $Z_L$, (not shown) on the top edge.
    (b) The XZZX code for $d=5$. All weight-4 stabilizers are equivalent, whereas the weight-2 boundary stabilizers are complementary to the adjacent weight-4 stabilizer. The logical $X_L$ used to define the equivalence classes is shown on the left edge, with a corresponding mixed $Z_L$ operator (not shown) on the top edge.}
    \label{fig:Surface code}
\end{figure}

As illustrated in \figref{fig:Surface code}, both codes consist of a square grid of $N=d^2$ qubits on which $N-1$ local (two- or four-qubit) stabilizers act. The stabilizers commute and square to identity, which means that they split the Hilbert space into size-2 blocks specified by the parity ($\pm 1$ eigenvalues) with respect to each stabilizer. The logical code space, $\ket{0}_L$ and $\ket{1}_L$, is taken to be the all even-parity block. Measuring the stabilizers projects the state onto one of the blocks and any odd-parity measurement is an indication that the state has been affected by errors and escaped the code space. The set of odd-parity measurements with their coordinates on the grid is referred to as the syndrome.   

%=================================================
%=================================================

Assuming Pauli noise, each qubit may be subject to one of three errors: $X$, $Z$, or $Y=XZ$ (up to an overall phase), occuring with rates $p_x$, $p_y$, and $p_z$, respectively, such that the total error rate is $p=p_x+p_y+p_z$, defined per measurement cycle of the syndrome.  %\cite{PhysRevA.72.052326}. 
This type of error model gives a standard, approximate, representation of qubit decay and dephasing~\cite{PhysRevA.88.012314, Tomita2014}. In addition, in practice, the measurement of the stabilizers require additional ancilla qubits that are entangled with the (data) qubits. As these ancillas are also noisy and the measurement circuit involves several two-qubit gates, there will be errors in the syndrome readout, which complicates the decoding significantly~\cite{wang2009threshold, PhysRevA.89.022321}. Also, in standard qubit architectures, `leakage' out of the qubit space leads to additional complications, as do possible correlated errors affecting two or more qubits~\cite{PhysRevA.81.022317, PhysRevA.88.042308, o2017density}.  
Nevertheless, to benchmark and discuss the basic features of the decoder we will use the basic error model and assume perfect syndromes. 

Logical operators act within the code space, thus commuting with the stabilizers. For the surface code they are a string of $Z$'s between two edges ($Z_L$) and a string of $X$'s between the other two edges ($X_L$). For XZZX they are in general mixed operators, except on the diagonals.

An arbitrary configuration of errors, or error chain, $C\in\{I,X,Y,Z\}^{\otimes N}$, is specified by an element in $\{I,X,Y,Z\}$ acting on each qubit, where $I$ is the identity operator. Each $C$ has a number of non-identity operators, quantified by $n_x$, $n_y$, and $n_z$, and gives a unique syndrome $s(C)$. It is a mapping many to one: a chain gives a unique syndrome, but a syndrome does not give a unique chain. The a priori probability of chain $C$, i.e., before measuring a syndrome, is given by
\begin{equation}
\label{eqn:chainprob}
    \pi_C = (1-p)^N \pi^r_C\,,
\end{equation}
where
\begin{equation}
   \pi^r_C = \mleft( \frac{p_z}{1-p} \mright)^{n_z} \mleft( \frac{p_x}{1-p} \mright)^{n_x} \mleft( \frac{p_y}{1-p} \mright)^{n_y} 
   \label{eq:pi_rel}
\end{equation}
is the probability relative to the empty chain.

For an arbitrary syndrome $s$, consider an arbitrary chain $C$ consistent with that syndrome. The equivalence class $E(C)$ of the chain $C$ is the set of all $2^{N-1}$ chains generated by acting with all elements of the stabilizer group. (A general element of the stabilizer group consists of a product of stabilizers or the identity operator acting on the $N-1$ faces of the code.)  %$E=\{f=f^*g:  g\in G\}$. ($G$ is stabilizer group. Clean up. Not sure how formal we want to be.) 
Three other equivalence classes with an equal number of chains are generated by acting with one of the logical operators $X_L$, $Z_L$, or $Y_L = X_L Z_L$. The equivalence class of the chain $E\in \{1,X,Y,Z\}$ is then decided by whether it commutes with $X_L$ and/or $Z_L$~\footnote{In order to uniquely specify the equivalence class, it is necessary to choose a specific representation of the logical operators. As the error chain does not commute with all stabilizers, deforming a logical operator may change the equivalence class classification. For the surface code we choose $X_L$ as the product of $X$ operators on the left edge and $Z_L$ as the product of $Z$ operators on the top edge, such that the class of the chain can be identified from the the parity of $Z$ errors on the left edge and the parity of $X$ errors on the top edge. Similarly for the XZZX code, where the operators on the edges are mixed.}.

The product of any two chains with the same syndrome gives a chain with a trivial syndrome, but, unless the two chains are in the same equivalence class, the product will correspond to a logical operator. Thus, maximum-likelihood decoding corresponds to providing an arbitrary correction chain in the most likely equivalence class. 

%%%%%%%%%%%%%%%%%%%%%%%%%%%%%%%%%%%%%%%%%%%%%%%%%%%%%%%%%%%%%%%

\section{Model and decoder}
\label{model}

%%%%%%%%%%%%%%%%%%%%%%%%%%%%%%%%%%%%%%%%%%%%

\subsection{Effective weight parametrization}

As discussed in \secref{background}, we only consider single-qubit Pauli errors with error rates $p_x+p_y+p_z=p$ and assume noise-free stabilizer measurements. We also assume $Z$-biased noise $p_z \geq p_x,p_y$. After an ideal stabilizer measurement cycle, the probability of a particular error chain with $n_x$, $n_y$, and $n_z$ errors is (relative to the empty chain) given by \eqref{eq:pi_rel}. %$\pi^r=(\frac{p_z}{1-p})^{n_z}(\frac{p_x}{1-p})^{n_x}(\frac{p_y}{1-p})^{n_y}$. 
It is convenient to rewrite this in terms of the most likely error, using the relative error rate $\tilde{p}_z\equiv \frac{p_z}{1-p}$, as
\begin{equation}
\pi^r = \mleft( \tilde{p}_z \mright)^{w} = e^{-\beta w}\,,
\label{eq:pi_n}
\end{equation}
with $\beta = - \ln{\tilde{p}_z}$ taking the role of an inverse temperature. 
Here we have defined the {\em effective weight} of the chain as
\begin{equation}
w = n_z + \alpha_x n_x + \alpha_y n_y \,, 
\end{equation}
with $\alpha_i$ given by
\begin{eqnarray}
\mleft( \frac{p_z}{1-p} \mright)^{\alpha_x} &=& \mleft( \frac{p_x}{1-p} \mright) \\
\mleft( \frac{p_z}{1-p} \mright)^{\alpha_y} &=& \mleft( \frac{p_y}{1-p} \mright)\,.
\label{eq:alpha_def}
\end{eqnarray}
The set of parameters $(\tilde{p}_z,\alpha_x,\alpha_y)$ provides a convenient parametrization of the 
space of error rates $(p_x,p_y,p_z)$, with $p_x+p_y+p_z=p\leq 1$. For fixed $\alpha_x$ and $\alpha_y$, which includes standard error models such as depolarizing, uncorrelated, and pure phase- or bit-flip noise, the single scalar $w$ specifies the probability of an error chain according to \eqref{eq:pi_n}. For this reason, the representation is also used for enhanced minimum-weight-matching decoders, where the weight of edges correspond to their relative probability~\cite{criger2018multi, bonilla2020xzzx}.

Limiting to $p_x=p_y$ with $\alpha\equiv \alpha_x=\alpha_y$, the total error rate $p$ is related to $\tilde{p}_z$ by $\tilde{p}=\tilde{p}_z+2\tilde{p}^\alpha_z$, with $\tilde{p}\equiv \frac{p}{1-p}$. Depolarizing noise corresponds to $\alpha=1$, such that $w=n_x+n_y+n_z$, while pure phase-flip noise (for $p\leq 0.5$) corresponds to $\alpha=\infty$, such that any chain with an $X$ or $Y$ error would have infinite weight, consistent with their vanishing probability. The parametrization for $Z$-biased noise in \cite{PhysRevLett.120.050505, PhysRevX.9.041031, bonilla2020xzzx} in terms of fixed $\eta=\frac{p_z}{p_x+p_y}$ and $p_x=p_y$ corresponds to fixed $\alpha$ only for depolarizing noise $\eta=\frac{1}{2}$ and pure phase-flip noise $\eta=\infty$, while in general there is a mapping
\begin{equation}
\alpha = \frac{\ln{\tilde{p}_z}-\ln{2\eta}}{\ln{\tilde{p}_z}} \,.
\end{equation}
Which parametrization to use for a single bias and error rate is a matter of convenience, but the advantage of the $\alpha$ parametrization is that the weight of a chain is conserved under changes of the overall error rate. (A disadvantage could be that the mapping between total and individual error rates is less direct.) In practice, this means that for a given syndrome, the same `lightest' chains will most likely be independent of error rate. 

For reference, ``uncorrelated'' noise, i.e., independent bit-flip and phase-flip errors, would correspond to $\alpha_x=1$ and $\alpha_y=2$. Note, however, that for uncorrelated noise, $p$ is conventionally defined as the independent rate of $X$ or $Z$ errors. The definition of effective length should still reflect the relative error rates, such that \eqref{eq:alpha_def} is replaced by
\begin{equation}
\mleft[ \frac{p(1-p)}{(1-p)^2} \mright]^{\alpha_y}=\frac{p^2}{(1-p)^2} \,.
\end{equation}
%

%%%%%%%%%%%%%%%%%%%%%%%%%%%%%%%%%%%%%%%%%%%%

\subsection{Effective weight and degeneracy decoder}

%%%%%%%%%%%%%%%%%%%%%%%%%%%%%%

\subsubsection{Calculating the probabilities of equivalence classes}

The probability of an equivalence class $E$ of a given syndrome $s$ is simply the probability of all chains in the class normalized with respect to all chains that are consistent with the syndrome:

\begin{equation}
\label{eqn:bolzman_sum}
    P_{E} = \frac{\sum_{C\in E} \pi_C}{\sum_{C:s(C)=s}\pi_C} = \frac{\sum_{w} N_{E}(w)\, e^{-\beta w}}{\sum_{E',w} N_{E'}(w)\, e^{-\beta w}}\,,
\end{equation}
where $N_E(w)$ is the total number of unique chains in class $E$ with effective weight $w$.
The numbers  $N_E(w)$ are combinatorial; they depend on the syndrome, but are independent of the overall error rate given by $\beta$, provided the relative error rates set by $\alpha$ are fixed. In principle, $N_E(w)$ can be calculated by finding one chain, using, e.g., minimum weight matching, acting on this chain with the full stabilizer group and, to switch equivalence class, the logical operators.

Since the number of elements in the stabilizer group grows exponentially with the number of qubits $N$, it quickly becomes infeasible to find or store the complete set of values. However, in the case that the partition function $Z_E$ of each equivalence class is dominated by the contribution of the lightest chains, we expect to obtain a good approximation by having an explicit count of only those chains:
\begin{equation}
    Z_E = \sum_{w} N_{E}(w)\, e^{-\beta w}\approx N_{E}^*\, e^{-\beta w^*_E} \,,
    \label{eqn:ZE_approx}
\end{equation}
where $w^*_E$ is the effective weight of lightest chain(s) in the class and $N_E^*=N_{E}(w^*_E)$ is the number of such chains. Thus, the probability of an equivalence class $P_E=Z_E/\sum_{E'}Z_{E'}$ 
according to this approximation requires knowledge of an octet of values
\begin{equation}
\mleft\{ w^*_E, N_{E}^* \mright\}_{E = \{1, X, Z, Y\}}\,. 
\end{equation}
%

%%%%%%%%%%%%%%%%%%%%%%%%%%%%%%

\subsubsection{The algorithm}

The EWD algorithm, which uses the above approximation, is outlined in pseudo-code form in Algorithm~\ref{alg:ewd_decoder}, using a hash table to store and identify unique chains. The Metropolis update rule (for more detail, see \secref{discussion}) that sets the probability of accepting a new error chain is operated at an error rate $p_{\text{sample}}$, which may be different from the physical error rate $p$. Since the weight of an error chain is set by the bias $\alpha$, independently of $p$, chains can be generated and stored at an arbitrary error rate. Sampling chains at an appropriately chosen (typically high) error rate has the advantage of avoiding that the algorithm becomes stuck in a low-weight configuration. The sampling error rate is effectively a hyperparameter of the decoder algorithm, which should be optimized such that the logical failure rate is minimized. We remark that in benchmarking the decoder using randomly generated error chains, it is of course important that the starting chains in the four equivalence classes are well randomized to ensure that it is not possible to underestimate the failure rate.

%=================================================
\begin{algorithm}
\SetAlgoNoLine
\SetInd{0em}{1em}
  \If{Syndrome $s$ is nontrivial}
  {\Indp
    \ForEach{equivalence class $E$}{
      Create empty hash table $T_E$\;
      Get an error chain $C \in E$ consistent with $s$\;
      Calculate the weight of $C$: $w$\;
      Calculate a hash of $C$: $H$\;
      Store $w$ in $T_E[H]$\;
      \For{$N$ iterations}{
        Apply a random stabilizer to $C$ to get a new chain $C'$\;
        Calculate the weight of $C'$: $w'$\;
        Use $w$ and $w'$ to calculate the relative probability $a = \pi_{C'} / \pi_{C}$ using the error rate $p_{\text{sample}}$\;
        Generate a uniform random number $u \sim \mathop{U}(0, 1)$\;
        \If{$a > u$}{
          $C \gets C'$ \; %\\
          $w \gets w'$ \; %\\
          Calculate a hash of $C$: $H$\;
          Store $w$ in $T_E[H]$\;
        }
      }
      Find lowest weight $w_E^*$ in $T_E$ and count its multiplicity $N_E^*$\;
    }
    Output probability of each equivalence class according to $P_E=Z_E/\sum_{E'}Z_{E'}$ with $Z_E=N_{E}^*\, e^{-\beta w^*_E}$ using the physical error rate $p$\;
    Apply an arbitrary correction chain from the equivalence class with the highest $P_E$ on the physical qubits\;
  }
\caption{The EWD decoder}
\label{alg:ewd_decoder}
\end{algorithm}
%=================================================

A variation of the EWD algorithm, referred to as ``All" in \secref{results}, evaluates the probability based on all unique observed chains, i.e., with $Z_E = \sum_{w} N_{E}(w)\, e^{-\beta w}$. If keeping the information content per syndrome to a minimum is not a priority, this version of the algorithm is more accurate, at least for non-integer $\alpha$.

%%%%%%%%%%%%%%%%%%%%%%%%%%%%%%

\subsubsection{Limitations of applicability}

Formally, the requirement for the approximation of \eqref{eqn:ZE_approx} to hold is that the probability density for chains of length $w$, $N_E(w)e^{-\beta w}$, is monotonically decreasing with chain weight. Otherwise the probability of each equivalence class will not be dominated by the lightest chains, which invalidates the basic assumption of the decoder. In other words, it should satisfy
\begin{equation}
N_{E}(w+\delta w)\, e^{-\beta\delta w} < N_{E}(w)
\end{equation}
for any $w\geq w^*_E$, with $\delta w$ the shortest difference between effective weights. Correspondingly, for large $w$, $\frac{d\ln{N_E}}{dw}\lesssim \beta$. Generally, the number of chains grows exponentially with the length (for $w\ll d^2$) as (averaged over syndromes and equivalence classes) $\sum_EN_E\sim d^{aw}$, where we have found that the parameter $a>0$ depends on type of code and error model, but only weakly on error rate. The latter dependency follows from the changes in the distribution of syndromes with error rate. This gives a corresponding upper bound on $d$ versus error rate, as $d\lesssim e^{\beta/a}$ for the approximation of estimating the partition function using only the lightest chains.

In practice, going beyond this limit to larger code distances or larger error rates (smaller $\beta$), the EWD decoder will perform worse than optimal. In this regime, the distribution $N_{E}(w)\, e^{-\beta w}$ will be centered on heavier chains, such that evaluating the partition function based on the lightest chains will be an increasingly poor approximation with increasing $d$ or increasing $p$.  We have not explored the limitations in detail, but the relatively sharp threshold presented for depolarizing noise (see \secref{results}) indicates that the decoder gives close to optimal results for $d<20$. For sufficiently low error rates (depending on $d$), the decoder will be optimal, as the lightest chains will dominate asymptotically. 

For general non-integer $\alpha$, it is less clear how to motivate the approximation, as $\delta w$, the difference between effective weights, can become arbitrarily small. Nevertheless, as discussed in \secref{results} below, even though the approximation may be quantitatively quite poor for specific syndromes,  aggregated over many syndromes, the effect of this discrepancy is found to be small.

%%%%%%%%%%%%%%%%%%%%%%%%%%%%%%%%%%%%%%%%%%%%%%%%%%%%%%%%%%%%%%%

\section{Results}
\label{results}

%%%%%%%%%%%%%%%%%%%%%%%%%%%%%%%%%%%%%%%%%%%%

\subsection{Test cases}

To demonstrate the capacity of the EWD decoder we apply it to several test cases: depolarizing noise ($\alpha=1$), moderately biased noise ($\alpha=5$), and highly biased noise ($\alpha= \text{10,000}$), focusing primarily on the XZZX code. For all these cases, the Metropolis sampling is done at a high error rate ($p_{\text{sample}} = 0.3$) compared to the error rate at which the syndromes are generated. The decoder is run for $25d^5$ Metropolis steps in each equivalence class. Due to the long correlation length, only every 5th step is considered and added to the set of observed chains throughout all tests. Unless otherwise stated, equivalence classes are seeded with a random chain consistent with that syndrome and class. For each configuration of $d$ and $p$, syndromes are randomly generated and solved in order to estimate the logical failure rate $P_f$. The latter is the fraction of all generated syndromes for which the randomly generated chain that provides the syndrome is not in the most likely equivalence class predicted by the decoder for that syndrome. For $d\leq 11$, 10,000 syndromes are solved; for $d=13$, at least 4,000; for $d=15$, at least 2,000. Error bars indicate one standard deviation based on the number of sampled syndromes and the mean logical failure rate. 

\begin{figure}

    \begin{tikzpicture}
    \node[inner sep=0pt] (y) at (0,0)
    {\includegraphics[width=\linewidth]{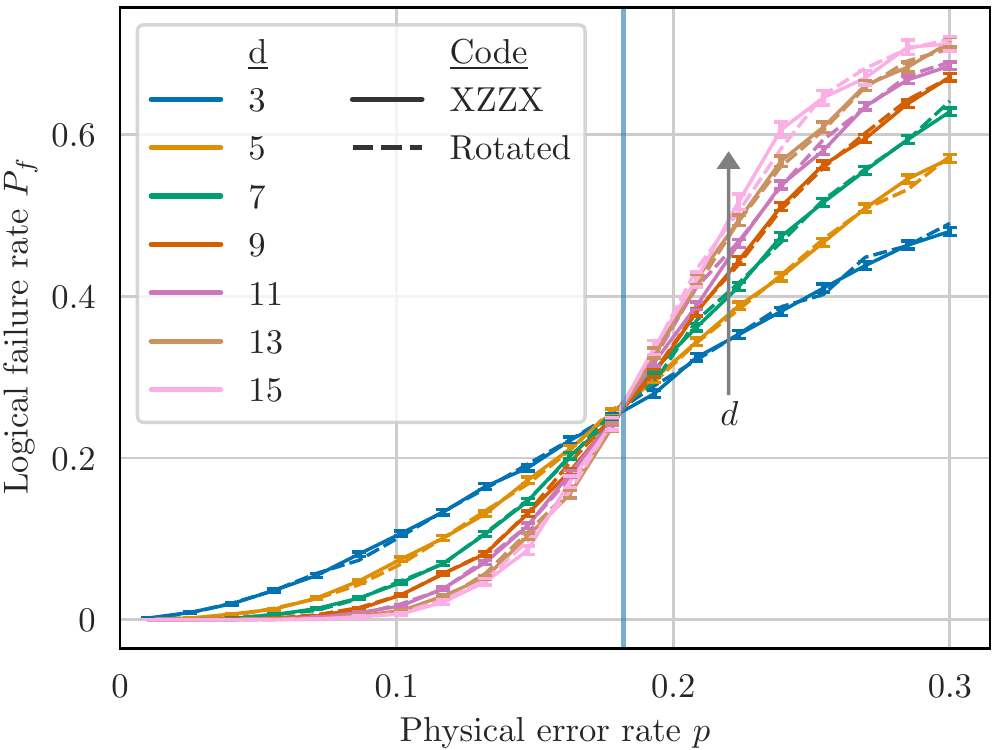}};
    \node[circle, draw=white, draw opacity=0, text=black, scale=1.2] (a) at (-3.9, 3.3) {(a)};
    \end{tikzpicture}
    
    \begin{tikzpicture}
    \node[inner sep=0pt] (y) at (0,0)
    {\includegraphics[width=\linewidth]{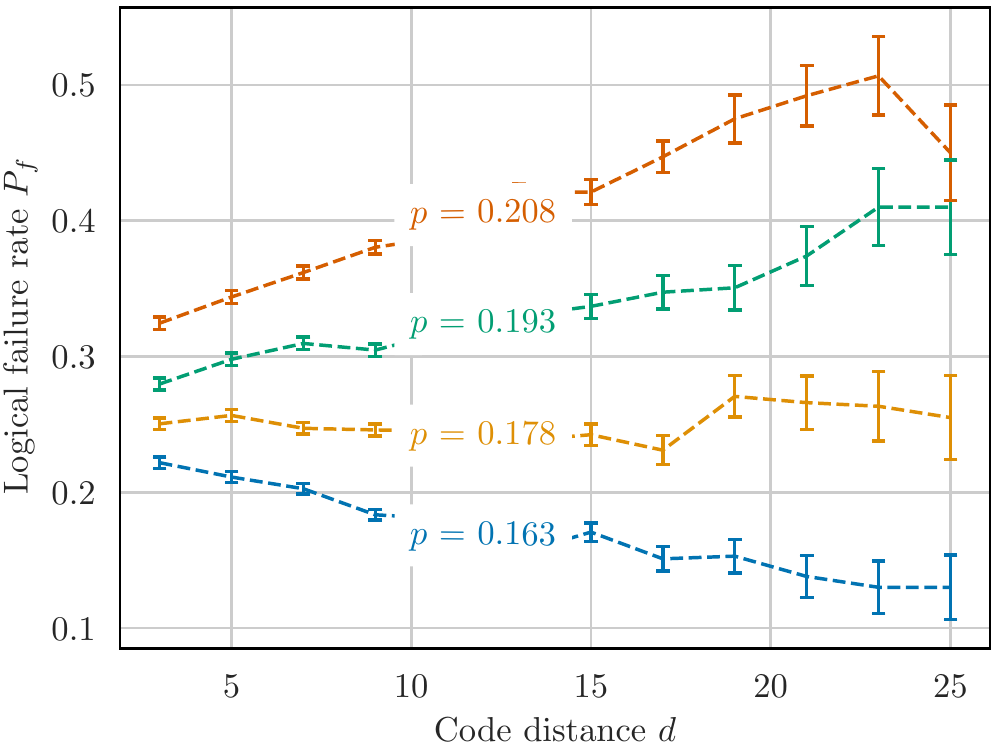}};
    \node[circle, draw=white, draw opacity=0, text=black, scale=1.2] (b) at (-3.9, 3.3) {(b)};
    \end{tikzpicture}
    
    \caption{Performance of the EWD decoder for depolarizing noise.
    (a) Logical failure rate $P_f$ as a function of physical error rate $p$ for depolarizing noise ($\alpha=1$) for the XZZX code and the rotated surface code using the EWD decoder on several different code distances $d$. The vertical blue line marks the approximate threshold at \unit[18.2]{\%}.
    (b) Logical failure rate as a function of code distance for error rates close to the threshold for depolarizing noise on the XZZX code. The transition from decreasing to increasing failure rate with increasing code distance indicates a threshold near an error rate $p \approx \unit[18]{\%}$. The number of syndromes for $d>15$ is at least 200.}
    \label{fig:xzzx_alpha1}
\end{figure}

\begin{figure}

    \includegraphics[trim=0cm 0cm 0cm 0cm, clip=true, width=0.48\textwidth]{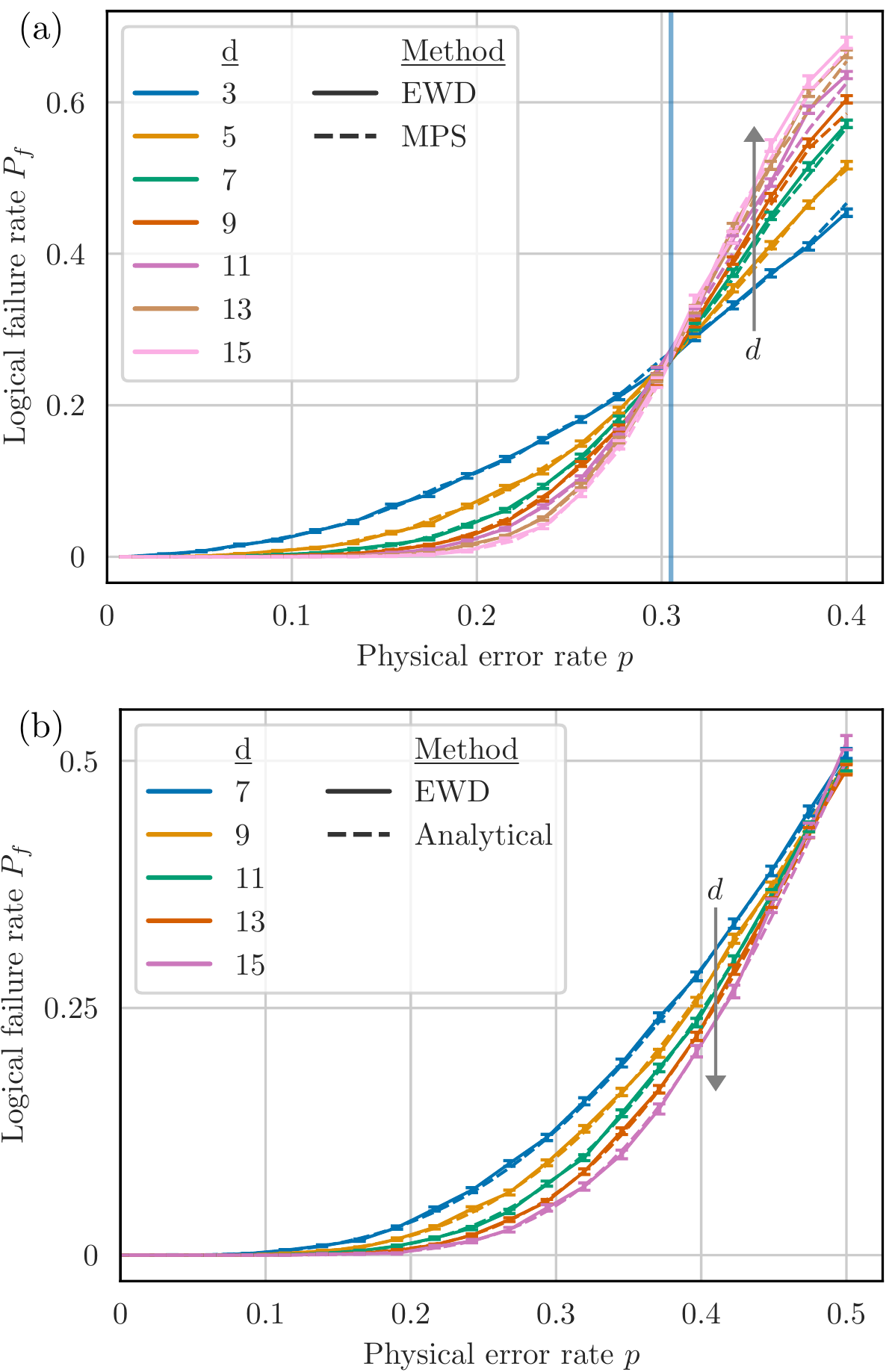}
    \caption{Performance of the EWD decoder for biased noise on the XZZX code.
    Logical failure rate $P_f$ as a function of physical error rate $p$ for  (a) moderately biased noise, $\alpha=5$, and (b) large bias $\alpha=10000$. In (a) compared to data obtained using the MPS decoder as implemented in~\cite{qecsim}. The vertical line indicates the threshold at \unit[30.5]{\%}. In (b) compared to exact results [\eqref{Eq:exact}] for pure phase-flip noise $\alpha=\infty$.}
    \label{fig:xzzx_bigalphas}
\end{figure}

As shown in \figref{fig:xzzx_alpha1}, we find an approximate threshold for depolarizing noise which is close to that of the MLDs for perfect syndrome measurements: $p_{\text{th}}\approx \unit[18.5]{\%}$~\cite{wootton2012, hutter2014efficient, Bravyi2014}. Note that for depolarizing noise, the XZZX code and the rotated surface code are equivalent. For larger bias, $\alpha = 5$, we find good correspondence with the results using the MPS decoder as implemented in Ref.~\cite{bonilla2020xzzx, qecsim}, with a threshold near $p\approx \unit[30]{\%}$, as shown in \figpanel{fig:xzzx_bigalphas}{a}. At $p=0.30$ and $\alpha=5$, corresponding to $\eta=18.3$, the hashing bound~\cite{PhysRevA.54.3824} is in fact $p_{\text{hash}}=0.307$, confirming the exceptional nature of the XZZX code.

%=================================================
%=================================================

%=================================================
%=================================================

For very large bias, $\alpha=$ 10,000, results are expected to be very close to those for $\alpha=\infty$, i.e., pure $Z$ noise, with a threshold of \unit[50]{\%}. For the latter, the following exact expression follows from the fact that there is only a single pure $Z$ operator (on one diagonal) that commutes with all stabilizers. It is not part of the stabilizer group, i.e., it is a logical operator, which implies that for every syndrome there are only two possible error chains in two different equivalence classes. Consequently, maximum-likelihood decoding implies that the decoder will suggest the lighter of the two and will always fail when the heavier chain occurs. For odd $d$ (and $p\leq 0.5$), the expression is
\begin{equation}
    P_{f, \alpha = \infty}(p) = \sum_{w = \lceil d/2 \rceil}^{d} \binom{d}{w} p^w (1-p)^{d-w}\,.
    \label{Eq:exact}
\end{equation}
Figure~\figpanelNoPrefix{fig:xzzx_bigalphas}{b} shows good correspondence between the numerical data from the EWD decoder and this expression. 

\begin{figure}
	\centering
    \includegraphics[width=\linewidth]{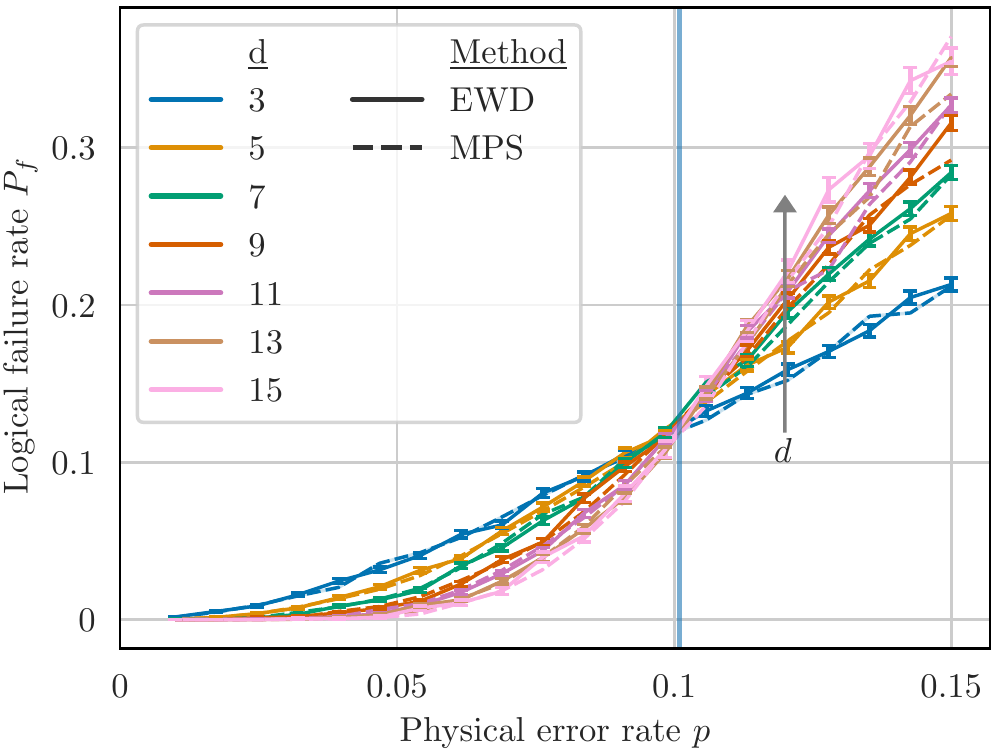}
	\caption{Performance of the EWD decoder for moderately biased noise, $\alpha=5$, on the rotated surface code. Logical failure rate as a function of physical error rate, obtained with the EWD  and MPS decoders for several code distances. The error rate used for sampling was $p_{\text{sample}}=0.3$ for all code distances except $d=\{13, 15\}$, which were sampled at $p_{\text{sample}}=0.15$. The vertical line indicates the approximate threshold at $p=0.101$.}
	\label{fig:rotated_alpha5}
\end{figure}

For the rotated surface code, we show in \figref{fig:xzzx_alpha1} results for depolarizing noise that, as expected, coincide with those for the XZZX code. In \figref{fig:rotated_alpha5}, we show results for $\alpha=5$, compared to the MPS decoder. As expected for phase-biased noise on the surface code, and contrary to the XZZX code, the threshold is reduced compared to depolarizing noise.

%=================================================
%=================================================

%%%%%%%%%%%%%%%%%%%%%%%%%%%%%%%%%%%%%%%%%%%%

\subsection{A specific syndrome example}

%%%%%%%%%%%%%%%%%%%%%%%%%%%%%%

\subsubsection{Integer weights}

\begin{figure}
    \begin{tikzpicture}
    \node[inner sep=0pt] (fig1) at (0,0)
      {\includegraphics[width=\linewidth]{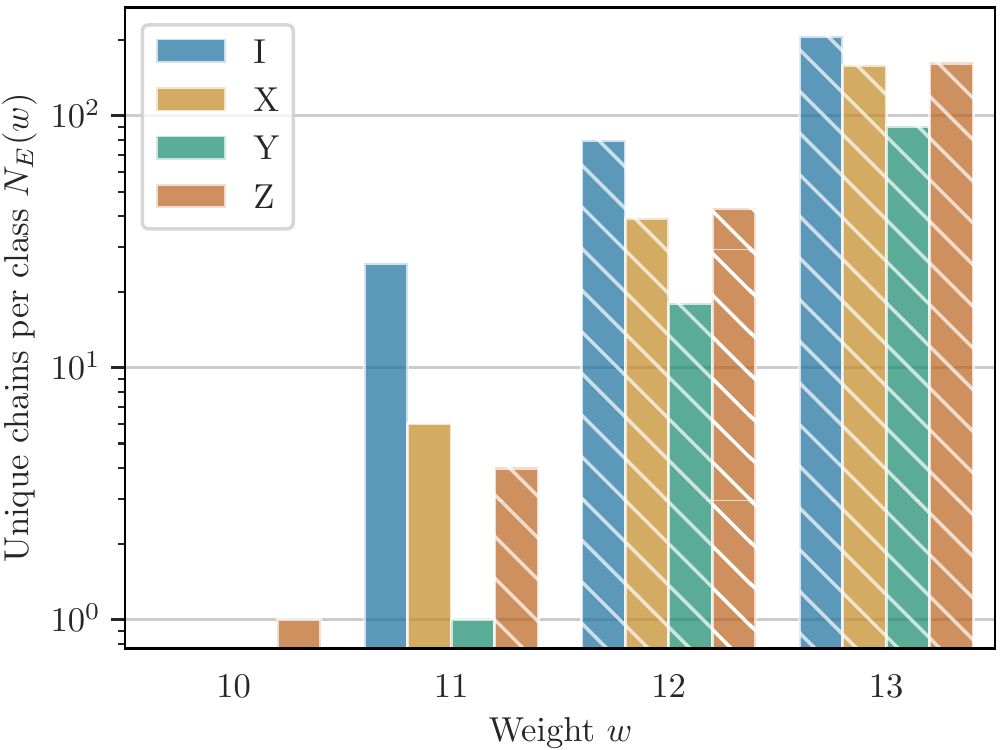}};
    \node[inner sep=0pt] (fig2) at (-0.55,2.1)
      {\includegraphics[width=0.23\linewidth]{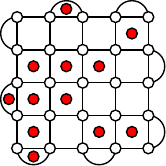}};
    \node[circle, draw=white, draw opacity=0, text=black, scale=1.2] (a) at (-3.9, 3.3) {(a)};
    \end{tikzpicture}
    
    \begin{tikzpicture}
    \node[inner sep=0pt] (y) at (0,0)
    {\includegraphics[width=\linewidth]{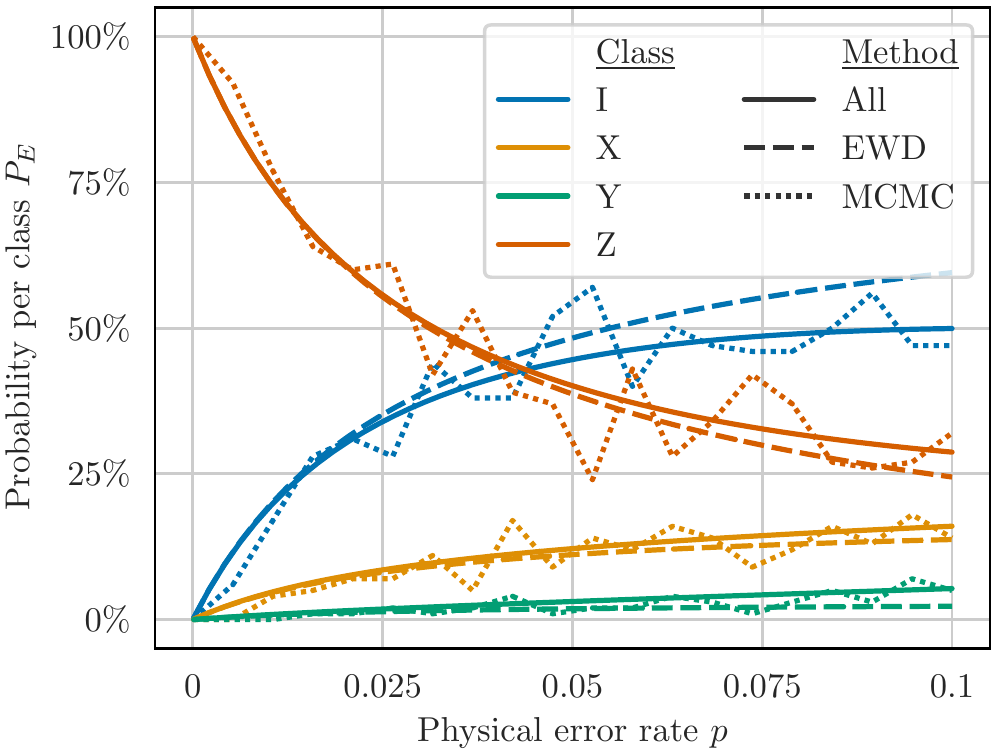}};
    \node[circle, draw=white, draw opacity=0, text=black, scale=1.2] (b) at (-3.9, 3.3) {(b)};
    \end{tikzpicture}

    % \subfloat[(b)]{
    % {\includegraphics[width=\linewidth]{special_syndrom_alpha2.pdf}}
    % }%[-0.5cm]
    \caption{Statistics of error chains obtained for a specific syndrome.
    (a) Number of unique error chains, $N_{E}(w)$, for biased noise $\alpha=2$, as a function of weight $w$ and equivalence class $E=\{$I,X,Y,Z\} for the syndrome on the $d=5$ XZZX code shown in the inset. The data is obtained from Metropolis sampling of error chains at an error rate of $p_{\text{sample}}=0.3$ for $25 \times 5^5$ Metropolis steps. Non-striped bars show the lowest weights in each equivalence class. Chains with $w>13$ are not shown.
    (b) Probability of equivalence class as a function of error rate for the syndrome in (a). Two different methods are used to evaluate \eqref{eqn:bolzman_sum} from the observed unique chains $N_{E}(w)$: using only the most likely chains in each equivalence class (EWD), and using all observed chains (All). Compared to the MCMC decoder, which is not based on identifying unique chains.}
    \label{fig:ex_alpha2}
\end{figure}

To better illustrate the workings and performance of the EWD decoder, it is useful to not only consider aggregate results on the logical failure rates over many syndromes, but to also look at individual syndromes. The example syndrome shown in the inset of \figpanel{fig:ex_alpha2}{a} has the interesting property that the most likely equivalence class may (depending on the noise bias) change as the error rate is increased. We consider a noise bias $\alpha=2$, where the weight of an error chain depends on the number of $X$, $Y$, and $Z$ errors according to $w = n_z + 2 (n_x + n_y)$.

Figure~\figpanelNoPrefix{fig:ex_alpha2}{a} shows the number of unique chains observed through Metropolis sampling at an error rate $p_{\text{sample}}=0.3$ for $25 \times 5^5$ Metropolis steps in each equivalence class starting from a randomized initial chain in each class. There is a single minimum-weight chain in class $Z$ ($w^*_Z=10, N_Z^*=1$), which will dominate the probability distribution at low error rates, while at larger error rates the larger number of one unit heavier chains in class $I$ ($w^*_I=11, N_I^*=12$) will be more important to decide most likely equivalence class, as seen in \figpanel{fig:ex_alpha2}{b}. Here we also see that using only the most likely chains in each class gives results that are in close correspondence with those given by evaluating $P_E$ based on all unique observed chains, as well as by using the MCMC decoder of Ref.~\cite{wootton2012}.

As is clear from \figpanel{fig:ex_alpha2}{b}, the data allows for the evaluation of the equivalence class probabilities for any error rate $p$. In this sense, the decoder is ``error-rate agnostic''. Notably, whereas the EWD curves are smooth (being generated from the same data), the MCMC results have large variations with error rate. For the latter, each error rate corresponds to generating a different Markov chain, and even though convergence criteria are applied on the most likely class, we find substantial variations in class probabilities. Even aggregated over 10,000 syndromes, the fluctuations are visible.

%=================================================
%=================================================

%%%%%%%%%%%%%%%%%%%%%%%%%%%%%%

\subsubsection{Non-integer weights}

\begin{figure}
    
    \begin{tikzpicture}
    \node[inner sep=0pt] (y) at (0,0)
    {\includegraphics[width=\linewidth]{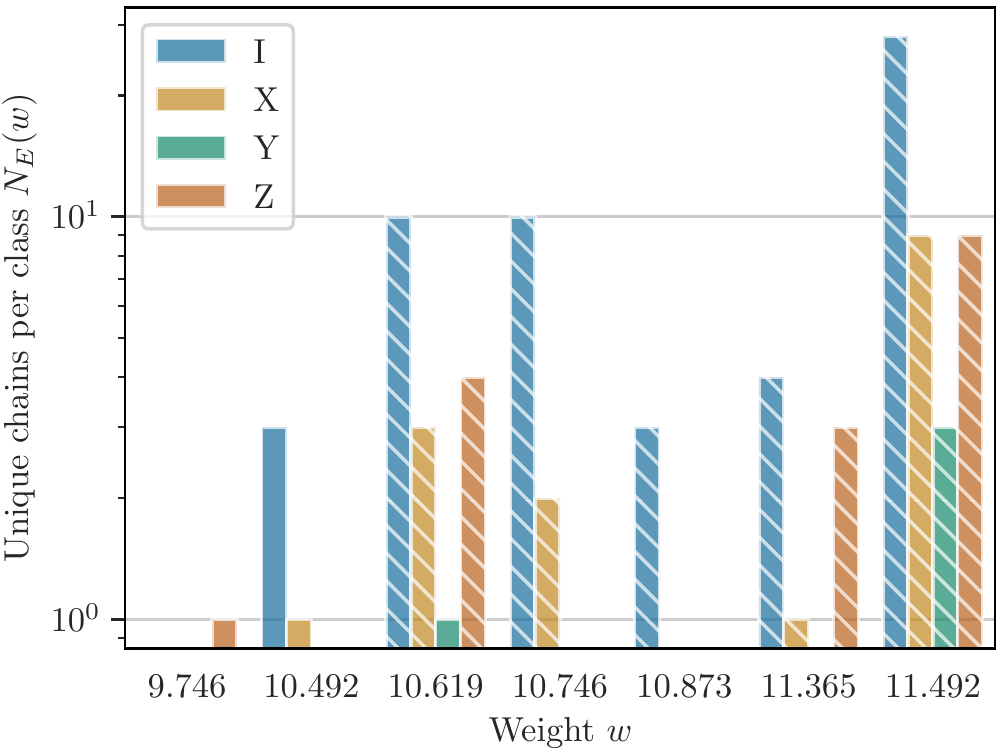}};
    \node[circle, draw=white, draw opacity=0, text=black, scale=1.2] (a) at (-3.9, 3.3) {(a)};
    \end{tikzpicture}%}\\[-0.5cm]
    
    %\subfloat[]{%
    \begin{tikzpicture}
    \node[inner sep=0pt] (y) at (0,0)
    {\includegraphics[width=\linewidth]{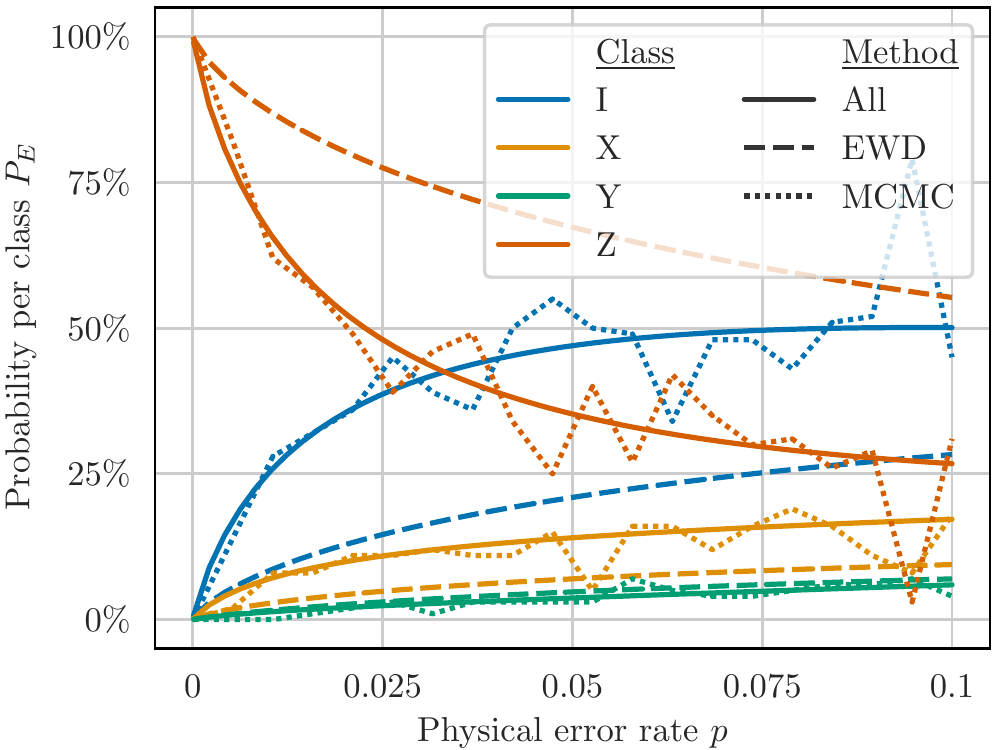}};
    \node[circle, draw=white, draw opacity=0, text=black, scale=1.2] (b) at (-3.9, 3.3) {(b)};
    \end{tikzpicture}%}\\[-0.5cm]
    
    \caption{Statistics of error chains obtained for the syndrome in the inset of \figpanel{fig:ex_alpha2}{a} for a non-integer noise bias.
    (a) Number of unique error chains, $N_{E}(w)$, for biased noise $\alpha=1.873$, as a function of weight $w$ and equivalence class $E=\{$I,X,Y,Z\}. Non-striped bars show the lowest weights in each equivalence class.
    (b) Probability of equivalence class as a function of error rate using data from (a). The methods are the same as in \figpanel{fig:ex_alpha2}{b}.}
    \label{fig:ex_alpha_off}
\end{figure}

The previous example considered biased noise with $\alpha=2$, i.e., an integer value, which implies that chain weights come in well-separated integer steps. Given limited knowledge of individual qubit noise biases, it may be reasonable to assume depolarizing noise or some other integer $\alpha$ bias, especially if the bias is large. Nevertheless, it is interesting to explore the consequences of non-integer $\alpha$. Figure~ \figpanelNoPrefix{fig:ex_alpha_off}{a} shows the distribution of unique chains for the same syndrome [inset of \figpanel{fig:ex_alpha2}{a}] for an arbitrary fraction, $\alpha=1.873$, close to $\alpha=2$. As expected, the bars of equal-weight chains are now fractionalized, i.e.\ fewer chains now have the same weight. 

\begin{figure}
	\centering
    \includegraphics[width=\linewidth]{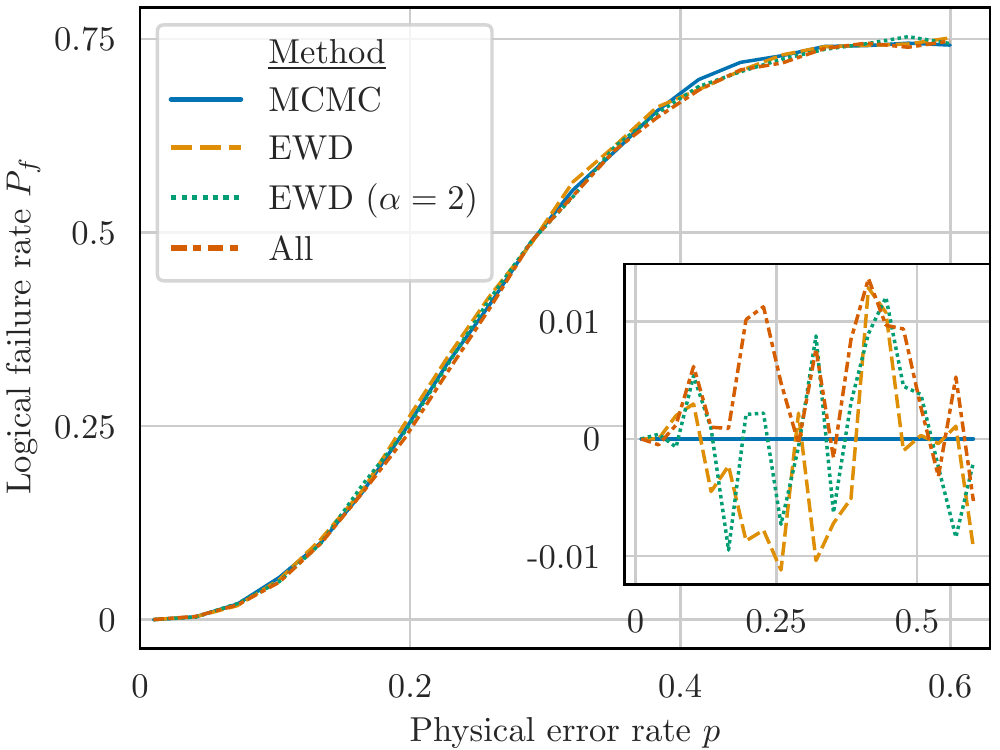}
	\caption{Comparison of decoding accuracy of different methods for non-integer bias $\alpha=1.873$ on the XZZX code. The methods used are the EWD decoder (i.e., accounting for only the lightest chains in each class) with $\alpha=1.873$ and $\alpha = 2$, weighing in all observed chains in each class (All), and the parallel-tempering (MCMC) decoder. The inset shows relative variations, which do not display any systematic differences. The data is obtained by averaging over 10,000 syndromes for $d=5$).}
	\label{fig:ex_alpha_off_aggregate}
\end{figure}

As shown in \figpanel{fig:ex_alpha_off}{b}, the EWD algorithm still gives the correct equivalence class for low error rates, and it qualitatively captures the transition between class $Z$ and class $I$, but at an incorrect error rate (outside of the view of the figure). In fact, for this syndrome, using only the most likely chains in each class, it is clearly a better approximation to round $\alpha$ to $\alpha=2$, as can be seen by comparing Figs.~\figpanelNoPrefix{fig:ex_alpha2}{b} and \figpanelNoPrefix{fig:ex_alpha_off}{b}. However, as seen from \figref{fig:ex_alpha_off_aggregate}, which shows the logical failure rate averaged over 10,000 randomly generated syndromes at $\alpha=1.873$, decoded using the EWD at $\alpha=1.873$ and $\alpha=2$, together with decoding using all identified chains, the difference at the aggregate level is small. The syndrome considered here as an example is thus an outlier. 

As argued in \secref{model}, it is only when it is important to have a compact representation of the equivalence class probabilities for a syndrome (such as for possible deep-learning purposes) that the EWD decoder using only the lowest-weight chains serves a potential purpose. If this is not the intent, there is no reason not to use all observed unique chains, regardless of weight, to evaluate the relative class probabilities using the first expression in \eqref{eqn:ZE_approx}.

%=================================================
%=================================================

%=================================================
%=================================================

%%%%%%%%%%%%%%%%%%%%%%%%%%%%%%%%%%%%%%%%%%%%

\subsection{Decoding efficiency}

The number of stabilizers grows exponentially with the code size $d$ as $2^{d^2}$, such that a brute force decoder aiming to generate all error chains of a given syndrome will have an exponential time complexity $t\sim O(2^{d^2})$. As discussed in Refs.~\cite{wootton2012, hutter2014efficient}, the time complexity of the MCMC decoder using parallel tempering is super-polynomial in $d$. It should be noted that this measure is based on convergence of the algorithm, implying close to optimal maximum-likelihood decoding. If the aim only is to improve on heuristic methods such as MWPM, the run time can scale significantly better with $d$. The ``single temperature'' (ST) algorithm formulated in Ref.~\cite{hutter2014efficient} is based on estimating the partition function \eqref{eqn:ZE_approx} in terms of the expectation value of the chain weight over a fixed number of Metropolis steps number in each class, $\langle w \rangle_{\beta,E}$. This was found to significantly improve on MWPM failure rates at the cost of an extra  $O(d^2)$ factor for the run time. 

\begin{figure}
	\centering
    \includegraphics[width=\linewidth]{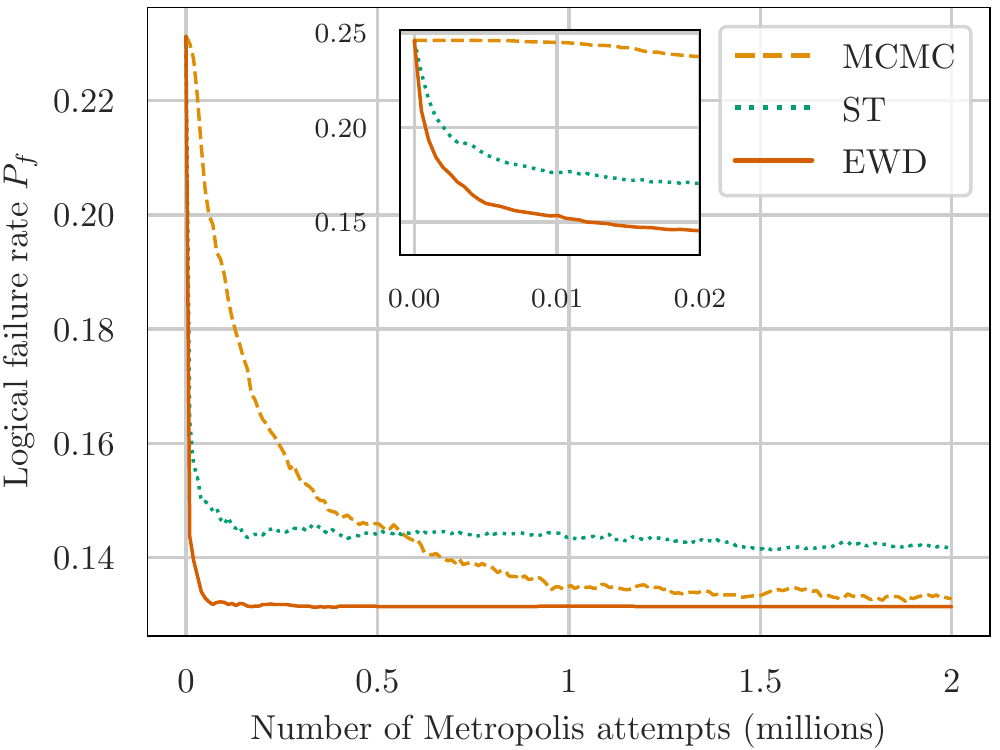}
	\caption{Convergence of the logical failure rate as a function of the number of Metropolis steps executed on a fixed set of $10,000$ syndromes for three different decoders: our EWD decoder, the MCMC decoder~\cite{wootton2012, hutter2014efficient}, and the ``single temperature'' (ST) Monte Carlo-based decoder of Ref.~\cite{hutter2014efficient}. The number of attempts is normalized with the number of equivalence classes (four) for the EWD and ST decoders, and with the number of parallel tempering levels (seven) for the MCMC decoder. All algorithms are seeded with the enhanced MWPM solutions~\cite{hutter2014efficient} based on Blossom V~\cite{kolmogorov2009blossom}, giving an initial logical failure rate of approximately $0.23$. The syndromes are generated on the standard (non-rotated) surface code with code distance $d=7$, subject to depolarizing noise ($\alpha=1$) with error rate $p=0.15$. The sampling error rate for the EWD decoder is set to $p_{\text{sample}}=0.25$. The inset shows the early convergence (using a different random set of 20,000 syndromes).}
	\label{fig:convergence}
\end{figure}

To study the time complexity of the EWD decoder we have considered two different regimes for depolarizing noise, of high error rates $p=0.15$ and asymptotically low error rates. For high error rates we compare the method to the MCMC decoder using parallel tempering, and to the aforementioned ST decoder. As shown in \figref{fig:convergence}, the convergence for moderate code sizes is significantly more rapid than the MCMC decoder, which we ascribe to the overhead of the latter from parallel tempering. The ST algorithm, although it has similar improved convergence as the EWD decoder, converges to a worse logical failure rate. In fact, the ST algorithm has a sub-optimal threshold for depolarizing noise (inset of Fig.~7 in Ref.~\cite{hutter2014efficient}) of between \unit[15]{\%} and \unit[16]{\%}. 

%=================================================
%=================================================

\begin{figure}
	\centering
    \includegraphics[width=\linewidth]{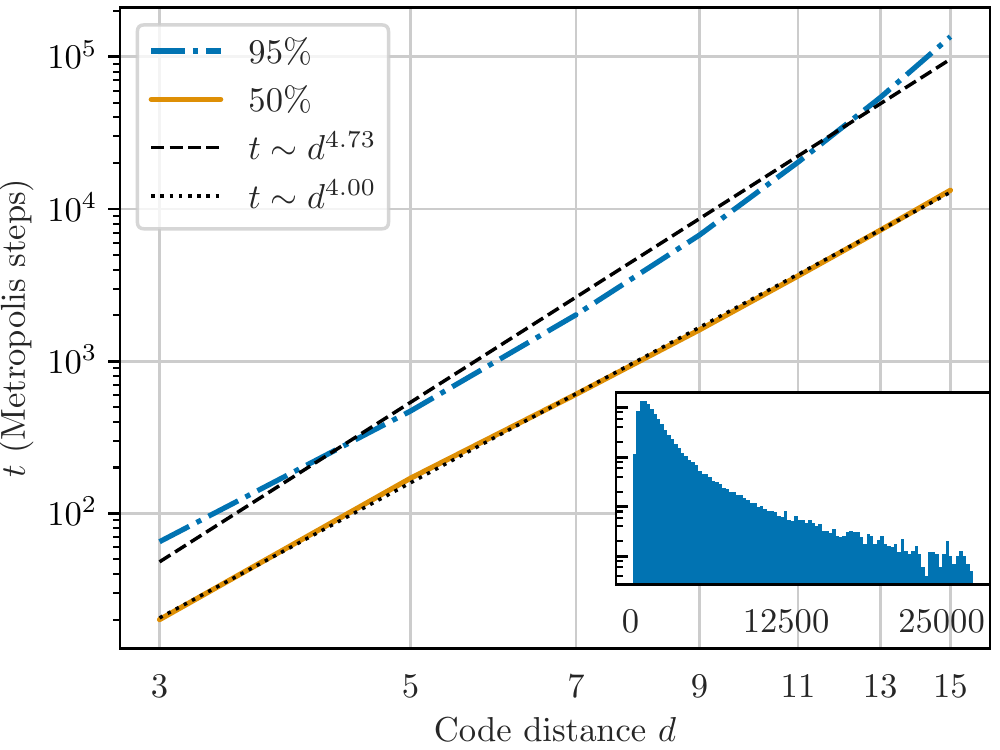}
	\caption{Estimated time complexity of the EWD decoder at asymptotically low error rate based on the number of Metropolis steps ($t$) required to identify a weight $w=(d-1)/2$ error chain for randomly generated syndromes that are guaranteed to have a chain with such weight. Shown is the maximum number of steps required to find 95\% and 50\% of the lowest weight chains and fitted to single exponents. The inset exemplifies the data for $d=9$ (containing $2\cdot 10^6$ syndromes) showing the number of syndromes versus the number of steps the decoder takes to find the weight $(d-1)/2$ chain.}
	\label{fig:time}
\end{figure}

For low error rates we have studied the time (in terms of the number of Metropolis steps) required to identify a weight $w_{\rm min}=(d-1)/2$ chain for syndromes for which these are the shortest possible chains. In the limit $p\rightarrow 0$, the shortest chains that can fail (given MLD) have weight $(d+1)/2$, so establishing the time required to find weight-$w_{\rm min}$ chains gives a good measure of the time complexity for low error rates. Specifically, we generate random error chains of weight $w=(d-1)/2$, deform them by a large number of random stabilizers and logical operators to produce heavy chains in all four equivalence classes, and then let the decoder find unique chains in the four classes, stopping when a weight-$w=(d-1)/2$ chain is identified. The results are plotted in \figref{fig:time}, showing an approximate scaling $t\lesssim O(d^5)$. However, there is an indication of superpolynomial behavior as $d$ is increased, which may be consistent with the complexity discussed in Ref.~\cite{wootton2012} for the MCMC decoder.

The inset in \figref{fig:time} gives the distribution of the number of steps required for $d=9$, as an example, showing that for most syndromes the lightest chain is easily found, while there is a long tail of rarer syndromes for which it takes the decoder significantly longer. The standard implementation of MWPM using the Blossom algorithm on a complete syndrome graph has a time complexity $O(d^6\log d)$, which can be reduced significantly in practice, for a realistic set of syndrome graphs, and by reducing the number of neighbors, to $O(d^2)$~\cite{higgott2021pymatching,PhysRevLett.108.180501}.

%=================================================
%=================================================

It is quite clear that the EWD decoder is not an efficient decoder compared to heuristic decoders such as MWPM and union find. Nevertheless, as discussed in \secref{sec:discussion-MWPM} below (see \tabref{tab:fail_frac}), since it is a maximum-likelihood decoder, the EWD decoder will provide lower logical failure rates.   

%%%%%%%%%%%%%%%%%%%%%%%%%%%%%%%%%%%%%%%%%%%%%%%%%%%%%%%%%%%%%%%

\section{Discussion}
\label{discussion}

%%%%%%%%%%%%%%%%%%%%%%%%%%%%%%%%%%%%%%%%%%%%

\subsection{Comparison with parallel-tempering-based decoders}

As the implementation of the EWD decoder uses Metropolis sampling of error chains similarly to the MCMC decoder formulated by \citet{wootton2012,hutter2014efficient}, a closer discussion of the similarities to and differences from these decoders is motivated. The probability of an equivalence class is the probability of an error chain in the class given a syndrome $s$ according to \eqref{eqn:bolzman_sum}. Using the Metropolis algorithm, a Markov chain of $M\ll 2^{N+1}$ (4 times the number of chains in each class) error chains $\{C_i\}$, $i=1,..,M$, representing the true distribution
\begin{equation}
P_C = \frac{\pi_C}{\sum_{C:s(C)=s}\pi_C} \,,
\end{equation}
is generated. The probability $P_{E}=\sum_{C\in E}P_C$ of the equivalence class $E$ is thus approximated by the relative number of observed chains in $E$:
\begin{equation}
P_{E}\approx \sum_{i:C_i\in E}1/M.
\end{equation}

The Markov chain is generated by acting with random stabilizers that deform the error chain $C\rightarrow C'$ without changing the syndrome, with the new chain accepted in standard fashion with probability $\text{min}(1,\pi_{C'}/\pi_C)$. However, stabilizers will not change the equivalence class, so in addition, it is also necessary to act with logical operators to generate chains in all equivalence classes. Whereas the stabilizers are weight-4 (or -2) operators, the logical operators are weight $d$, which makes these exponentially (with $d$) less likely to be accepted. A naive implementation would thus require exponential computational time to generate a Markov chain which is appropriately balanced between the equivalence classes.

In order to deal with this problem,  `parallel tempering'~\cite{earl2005parallel} is employed in \cite{wootton2012,hutter2014efficient}, in which chains are generated and exchanged between several layers of the stabilizer code with different error probabilities. The probability of accepting a swap of chains $C$ and $C'$ between layers with error rate $\vec{p}$ and $\vec{p}'$ is given by
\begin{equation}
\text{min} \mleft( 1, \frac{\pi_C(\vec{p}')\pi_{C'}(\vec{p})}{\pi_C(\vec{p})\pi_{C'}(\vec{p}')} \mright) \,.
\end{equation}
The error rates are progressed from the physical error rate to the top layer where all error chains are assumed equally likely, such that proposed logical operators are always accepted. For depolarizing noise $p_x=p_y=p_z=p/3$, the interpolation is straightforward, changing $p$ from the physical value to $p=0.75$ in the top layer~\cite{wootton2012}. For biased noise, we have implemented an interpolation at constant $\alpha$, with $\tilde{p}_z$ ranging from the physical value to $\tilde{p}_z=1$ in the top layer, where the effective weight is irrelevant~\cite{git}. An alternative approach to account for the heavy logical operators in the Monte Carlo sampling is to use Bennett's acceptance-ratio method to progress from high to low error rates, as discussed in Ref.~\cite{PhysRevA.88.062308}.

The accuracy of the MCMC decoder depends crucially on appropriately exploring all four equivalence classes through parallel tempering. It is thus quite sensitive to the convergence criteria. In particular, we have found that for low error rates the decoder has quite poor accuracy. We interpret this as being due to an inefficiency in finding the short-effective-length error chains that will dominate the class probabilities. At low error rates, i.e., low temperatures, the exploration effectively becomes frozen into a small non-representative part of the configuration space.  

In contrast, the EWD decoder is based on explicitly evaluating the contribution of the most likely chains. In order to find the chains, we use Metropolis sampling, but without parallel tempering. Since the algorithm is based on an explicit identification and count of unique chains, there is no need to find a properly class-balanced set of chains. In addition, even within a class, there is no need that the Markov chain actually follows the distribution corresponding to the physical error rate. As long as the effective weight of the chains is conserved, i.e., fixed $\alpha$, one may use a different error rate to explore and identify the chains, allowing for more efficient exploration.
%Given that the light error chains constitute a small fraction of all error chains this makes it feasible to explicitly track the number of unique short chains even for moderately large codes. 
From a computational perspective, the EWD algorithm can also be run efficiently within memory constraints by using a hash-table data structure for identifying chains as unique, as there is no need to store the actual chains, but only the total number of unique chains according to weight. 

%%%%%%%%%%%%%%%%%%%%%%%%%%%%%%%%%%%%%%%%%%%%

\subsection{Comparison with minimum-weight-matching algorithms}
\label{sec:discussion-MWPM}

The information used in the EWD decoder is the weight and number of the most likely error chains in each equivalence class. Standard minimum-weight-matching decoding has two severe shortcomings in this respect. First, bit- and phase-flip errors are treated as separate graph problems. This amounts to measuring the weight using a model of uncorrelated errors ($\alpha_z=\alpha_x=1$, $\alpha_y=2$), such that the suggested lowest-weight chain(s) may not be the actual lowest-weight chain(s). Second, degeneracy of chains with the same weight is not considered. The latter is crucial for the high performance of the EWD decoder, as without this information, syndromes are always decoded according to the $p\rightarrow 0$ class prediction. 

To quantify the effect of the first issue (incorrect weight of chains containing $Y$), we have studied the failure rate for chains of weight $(d+1)/2$ for depolarizing noise for the XZZX code (with equivalent results for the rotated surface code). These are the lightest error chains that will give failure, since for a small fraction of such chains there is a $(d-1)/2$-weight chain which is in another equivalence class. These chains will subsequently give the asymptotic failure rate at low $p$, $P_f\sim p^{(d+1)/2}$ with a prefactor that is given by the number of failing chains. Table \ref{tab:fail_frac} shows the results of this study. The MWPM algorithm fails on a significantly larger fraction of these chains, due to overestimating the weight of chains containing $Y$ errors.  

\begin{table}[h]
    \centering
    \begin{tabular}{ccc}
    \hline
    \hline
        d & MWPM & EWD\\
        \hline
        5 & 0.075 & 0.040 \\
         7 & 0.0086 & 0.0028 \\
          9 & 0.00073 & 0.00018 \\
    \hline     
    \end{tabular}
    \caption{Fraction of failed chains with weight $w=(d+1)/2$ for the XZZX code and depolarizing noise, comparing MWPM (using \cite{higgott2021pymatching}) and EWD. The relative fraction corresponds to the relative asymptotic ($p\rightarrow 0$) logical failure rate for the two decoders. Enough chains are sampled such that the digits specified are precise within one standard deviation.}
    \label{tab:fail_frac}
\end{table}

Using Metropolis sampling as in this work eliminates both the incorrect weight of chains and the failure to take into account degeneracy, but is computationally expensive. There is also a number of works which provide improvements to minimum-weight-matching schemes by reweighting graph edges to better reflect the actual error rates and account for degeneracy of chains. We note in particular the method to enhance MWPM presented in Ref.~\cite{criger2018multi}, which addresses both of these issues, using multiplicity counts and belief propagation on error rates, resulting in high thresholds for both uncorrelated and depolarizing noise on the surface code.

%%%%%%%%%%%%%%%%%%%%%%%%%%%%%%%%%%%%%%%%%%%%%%%%%%%%%%%%%%%%%%%

\section{Conclusion}
\label{conclusion}

We have implemented a decoding algorithm for topological stabilizer codes, which is based on evaluating the most likely equivalence class of error chains for a given syndrome, based on the effective weight and number of most likely chains in each class. This ``effective weight and degeneracy" (EWD) decoder is ``error-rate agnostic" in the sense of being invariant under changes that preserve the weight of error chains, corresponding to fixed relative bias of phase- to bit-flip errors. It thus represents a type of decoder which is intermediate between maximum-likelihood decoders, which work at a fixed error rate, and heuristic decoders, e.g., minimum weight matching, which are insensitive to the error rate.    

While related to the Markov-chain Monte Carlo-based decoder of Refs.~\cite{wootton2012, hutter2014efficient}, using the Metropolis algorithm to sample error chains, we find that the EWD decoder is significantly more efficient, flexible, and has better convergence properties. The reason for these advantages is that the error-chain sampling can be done independently in the equivalence classes and at a higher error rate than the physical, which facilitates efficient exploration. A drawback of the same freedom is that the results may be sensitive to the sampling error rate. We have found that it is useful to think of the sampling rate as a hyperparameter and tune it for best results. %Provided the decoder is only served properly disordered error chains in each equivalence class as starting configurations, it is not possible to underestimate the failure rate, up to statistical fluctuations. 
For optimal performance the decoder is limited to moderate code sizes or low error rates, for which the distribution of error chains is dominated by the most likely chains. However, for large code sizes the Metropolis sampling becomes prohibitively expensive, such that the limitations may not be very important for practical purposes.

We have found that the EWD decoder can be used as a relatively fast and flexible maximum-likelihood decoder for exploring novel stabilizer codes~\cite{srivastava2021xyz2}. 
In addition, we argue that the compact data representation, where, for a given error bias, each syndrome is described by a small set of numbers (weight and degeneracy of most likely chains in each class), can be suitable for supervised deep-learning approaches that would allow for fast decoding while having flexibility with respect to the overall error rate.

Future work should also further explore the inclusion of circuit level noise leading to imperfect stabilizer measurements~\cite{PhysRevA.83.020302, hutter2014efficient}, which can be incorporated into the algorithm by appropriately including the weight of faulty measurement events on a 3D lattice. Even though the EWD decoder is slow compared to matching or union-find decoders, and not practical for large codes, it may be useful for establishing approximate maximum-likelihood thresholds to provide upper bounds on code performance.

%%%%%%%%%%%%%%%%%%%%%%%%%%%%%%%%%%%%%%%%%%%%%%%%%%%%%%%%%%%%%%%

\begin{acknowledgments}

Computations were enabled by resources provided by the Swedish National Infrastructure for Computing (SNIC) and Chalmers Centre for Computational Science and Engineering (C3SE). We acknowledge the financial support from the Knut and Alice Wallenberg Foundation through the Wallenberg Centre for Quantum Technology (WACQT). The software for running the EWD and MCMC decoders is available on the repository~\cite{git}, while the implementation of the MPS decoder used in this work is available from Ref.~\cite{qecsim}.

\end{acknowledgments}

%%%%%%%%%%%%%%%%%%%%%%%%%%%%%%%%%%%%%%%%%%%%%%%%%%%%%%%%%%%%%%%

\bibliography{Bib}

%merlin.mbs apsrev4-1.bst 2010-07-25 4.21a (PWD, AO, DPC) hacked
%Control: key (0)
%Control: author (8) initials jnrlst
%Control: editor formatted (1) identically to author
%Control: production of article title (0) allowed
%Control: page (1) range
%Control: year (1) truncated
%Control: production of eprint (0) enabled
\begin{thebibliography}{86}%
\makeatletter
\providecommand \@ifxundefined [1]{%
 \@ifx{#1\undefined}
}%
\providecommand \@ifnum [1]{%
 \ifnum #1\expandafter \@firstoftwo
 \else \expandafter \@secondoftwo
 \fi
}%
\providecommand \@ifx [1]{%
 \ifx #1\expandafter \@firstoftwo
 \else \expandafter \@secondoftwo
 \fi
}%
\providecommand \natexlab [1]{#1}%
\providecommand \enquote  [1]{``#1''}%
\providecommand \bibnamefont  [1]{#1}%
\providecommand \bibfnamefont [1]{#1}%
\providecommand \citenamefont [1]{#1}%
\providecommand \href@noop [0]{\@secondoftwo}%
\providecommand \href [0]{\begingroup \@sanitize@url \@href}%
\providecommand \@href[1]{\@@startlink{#1}\@@href}%
\providecommand \@@href[1]{\endgroup#1\@@endlink}%
\providecommand \@sanitize@url [0]{\catcode `\\12\catcode `\$12\catcode
  `\&12\catcode `\#12\catcode `\^12\catcode `\_12\catcode `\%12\relax}%
\providecommand \@@startlink[1]{}%
\providecommand \@@endlink[0]{}%
\providecommand \url  [0]{\begingroup\@sanitize@url \@url }%
\providecommand \@url [1]{\endgroup\@href {#1}{\urlprefix }}%
\providecommand \urlprefix  [0]{URL }%
\providecommand \Eprint [0]{\href }%
\providecommand \doibase [0]{http://dx.doi.org/}%
\providecommand \selectlanguage [0]{\@gobble}%
\providecommand \bibinfo  [0]{\@secondoftwo}%
\providecommand \bibfield  [0]{\@secondoftwo}%
\providecommand \translation [1]{[#1]}%
\providecommand \BibitemOpen [0]{}%
\providecommand \bibitemStop [0]{}%
\providecommand \bibitemNoStop [0]{.\EOS\space}%
\providecommand \EOS [0]{\spacefactor3000\relax}%
\providecommand \BibitemShut  [1]{\csname bibitem#1\endcsname}%
\let\auto@bib@innerbib\@empty
%</preamble>
\bibitem [{\citenamefont {Shor}(1995)}]{PhysRevA.52.R2493}%
  \BibitemOpen
  \bibfield  {author} {\bibinfo {author} {\bibfnamefont {P.~W.}\ \bibnamefont
  {Shor}},\ }\bibfield  {title} {\enquote {\bibinfo {title} {Scheme for
  reducing decoherence in quantum computer memory},}\ }\href {\doibase
  10.1103/PhysRevA.52.R2493} {\bibfield  {journal} {\bibinfo  {journal}
  {Physical Review A}\ }\textbf {\bibinfo {volume} {52}},\ \bibinfo {pages}
  {R2493} (\bibinfo {year} {1995})}\BibitemShut {NoStop}%
\bibitem [{\citenamefont {Steane}(1996)}]{PhysRevLett.77.793}%
  \BibitemOpen
  \bibfield  {author} {\bibinfo {author} {\bibfnamefont {A.~M.}\ \bibnamefont
  {Steane}},\ }\bibfield  {title} {\enquote {\bibinfo {title} {{Error
  Correcting Codes in Quantum Theory}},}\ }\href {\doibase
  10.1103/PhysRevLett.77.793} {\bibfield  {journal} {\bibinfo  {journal}
  {Physical Review Letters}\ }\textbf {\bibinfo {volume} {77}},\ \bibinfo
  {pages} {793} (\bibinfo {year} {1996})}\BibitemShut {NoStop}%
\bibitem [{\citenamefont {Gottesman}(1997)}]{gottesman1997stabilizer}%
  \BibitemOpen
  \bibfield  {author} {\bibinfo {author} {\bibfnamefont {D.}~\bibnamefont
  {Gottesman}},\ }\href {https://arxiv.org/abs/quant-ph/9705052} {\emph
  {\bibinfo {title} {Stabilizer codes and quantum error correction}}}\
  (\bibinfo  {publisher} {California Institute of Technology},\ \bibinfo {year}
  {1997})\BibitemShut {NoStop}%
\bibitem [{\citenamefont {Nielsen}\ and\ \citenamefont
  {Chuang}(2000)}]{Nielsen2000}%
  \BibitemOpen
  \bibfield  {author} {\bibinfo {author} {\bibfnamefont {M.~A.}\ \bibnamefont
  {Nielsen}}\ and\ \bibinfo {author} {\bibfnamefont {I.~L.}\ \bibnamefont
  {Chuang}},\ }\href@noop {} {\emph {\bibinfo {title} {{Quantum Computation and
  Quantum Information}}}}\ (\bibinfo  {publisher} {Cambridge University
  Press},\ \bibinfo {year} {2000})\BibitemShut {NoStop}%
\bibitem [{\citenamefont {Terhal}(2015)}]{terhal2015quantum}%
  \BibitemOpen
  \bibfield  {author} {\bibinfo {author} {\bibfnamefont {B.~M.}\ \bibnamefont
  {Terhal}},\ }\bibfield  {title} {\enquote {\bibinfo {title} {Quantum error
  correction for quantum memories},}\ }\href {\doibase
  10.1103/RevModPhys.87.307} {\bibfield  {journal} {\bibinfo  {journal}
  {Reviews of Modern Physics}\ }\textbf {\bibinfo {volume} {87}},\ \bibinfo
  {pages} {307} (\bibinfo {year} {2015})}\BibitemShut {NoStop}%
\bibitem [{\citenamefont {Girvin}(2021)}]{Girvin2021}%
  \BibitemOpen
  \bibfield  {author} {\bibinfo {author} {\bibfnamefont {S.~M.}\ \bibnamefont
  {Girvin}},\ }\href@noop {} {\enquote {\bibinfo {title} {{Introduction to
  Quantum Error Correction and Fault Tolerance}},}\ } (\bibinfo {year}
  {2021}),\ \Eprint {http://arxiv.org/abs/2111.08894} {arXiv:2111.08894}
  \BibitemShut {NoStop}%
\bibitem [{\citenamefont {Kitaev}(2003)}]{kitaev2003fault}%
  \BibitemOpen
  \bibfield  {author} {\bibinfo {author} {\bibfnamefont {A.~Y.}\ \bibnamefont
  {Kitaev}},\ }\bibfield  {title} {\enquote {\bibinfo {title} {Fault-tolerant
  quantum computation by anyons},}\ }\href {\doibase
  10.1016/S0003-4916(02)00018-0} {\bibfield  {journal} {\bibinfo  {journal}
  {Annals of Physics}\ }\textbf {\bibinfo {volume} {303}},\ \bibinfo {pages}
  {2} (\bibinfo {year} {2003})}\BibitemShut {NoStop}%
\bibitem [{\citenamefont {Bravyi}\ and\ \citenamefont
  {Kitaev}(1998)}]{bravyi1998quantum}%
  \BibitemOpen
  \bibfield  {author} {\bibinfo {author} {\bibfnamefont {S.~B.}\ \bibnamefont
  {Bravyi}}\ and\ \bibinfo {author} {\bibfnamefont {A.~Y.}\ \bibnamefont
  {Kitaev}},\ }\href@noop {} {\enquote {\bibinfo {title} {Quantum codes on a
  lattice with boundary},}\ } (\bibinfo {year} {1998}),\ \Eprint
  {http://arxiv.org/abs/quant-ph/9811052} {arXiv:quant-ph/9811052} \BibitemShut
  {NoStop}%
\bibitem [{\citenamefont {Dennis}\ \emph {et~al.}(2002)\citenamefont {Dennis},
  \citenamefont {Kitaev}, \citenamefont {Landahl},\ and\ \citenamefont
  {Preskill}}]{dennis2002topological}%
  \BibitemOpen
  \bibfield  {author} {\bibinfo {author} {\bibfnamefont {E.}~\bibnamefont
  {Dennis}}, \bibinfo {author} {\bibfnamefont {A.}~\bibnamefont {Kitaev}},
  \bibinfo {author} {\bibfnamefont {A.}~\bibnamefont {Landahl}}, \ and\
  \bibinfo {author} {\bibfnamefont {J.}~\bibnamefont {Preskill}},\ }\bibfield
  {title} {\enquote {\bibinfo {title} {Topological quantum memory},}\ }\href
  {\doibase 10.1063/1.1499754} {\bibfield  {journal} {\bibinfo  {journal}
  {Journal of Mathematical Physics}\ }\textbf {\bibinfo {volume} {43}},\
  \bibinfo {pages} {4452} (\bibinfo {year} {2002})}\BibitemShut {NoStop}%
\bibitem [{\citenamefont {Raussendorf}\ and\ \citenamefont
  {Harrington}(2007)}]{Raussendorf2007}%
  \BibitemOpen
  \bibfield  {author} {\bibinfo {author} {\bibfnamefont {R.}~\bibnamefont
  {Raussendorf}}\ and\ \bibinfo {author} {\bibfnamefont {J.}~\bibnamefont
  {Harrington}},\ }\bibfield  {title} {\enquote {\bibinfo {title}
  {{Fault-Tolerant Quantum Computation with High Threshold in Two
  Dimensions}},}\ }\href {\doibase 10.1103/PhysRevLett.98.190504} {\bibfield
  {journal} {\bibinfo  {journal} {Physical Review Letters}\ }\textbf {\bibinfo
  {volume} {98}},\ \bibinfo {pages} {190504} (\bibinfo {year}
  {2007})}\BibitemShut {NoStop}%
\bibitem [{\citenamefont {Fowler}\ \emph
  {et~al.}(2012{\natexlab{a}})\citenamefont {Fowler}, \citenamefont
  {Mariantoni}, \citenamefont {Martinis},\ and\ \citenamefont
  {Cleland}}]{fowler2012surface}%
  \BibitemOpen
  \bibfield  {author} {\bibinfo {author} {\bibfnamefont {A.~G.}\ \bibnamefont
  {Fowler}}, \bibinfo {author} {\bibfnamefont {M.}~\bibnamefont {Mariantoni}},
  \bibinfo {author} {\bibfnamefont {J.~M.}\ \bibnamefont {Martinis}}, \ and\
  \bibinfo {author} {\bibfnamefont {A.~N.}\ \bibnamefont {Cleland}},\
  }\bibfield  {title} {\enquote {\bibinfo {title} {{Surface codes: Towards
  practical large-scale quantum computation}},}\ }\href {\doibase
  10.1103/PhysRevA.86.032324} {\bibfield  {journal} {\bibinfo  {journal}
  {Physical Review A}\ }\textbf {\bibinfo {volume} {86}},\ \bibinfo {pages}
  {032324} (\bibinfo {year} {2012}{\natexlab{a}})}\BibitemShut {NoStop}%
\bibitem [{\citenamefont {Kelly}\ \emph {et~al.}(2015)\citenamefont {Kelly},
  \citenamefont {Barends}, \citenamefont {Fowler}, \citenamefont {Megrant},
  \citenamefont {Jeffrey}, \citenamefont {White}, \citenamefont {Sank},
  \citenamefont {Mutus}, \citenamefont {Campbell}, \citenamefont {Chen},
  \citenamefont {Chen}, \citenamefont {Chiaro}, \citenamefont {Dunsworth},
  \citenamefont {Hoi}, \citenamefont {Neill}, \citenamefont {O'Malley},
  \citenamefont {Quintana}, \citenamefont {Roushan}, \citenamefont
  {Vainsencher}, \citenamefont {Wenner}, \citenamefont {Cleland},\ and\
  \citenamefont {Martinis}}]{Kelly2015}%
  \BibitemOpen
  \bibfield  {author} {\bibinfo {author} {\bibfnamefont {J.}~\bibnamefont
  {Kelly}}, \bibinfo {author} {\bibfnamefont {R.}~\bibnamefont {Barends}},
  \bibinfo {author} {\bibfnamefont {A.~G.}\ \bibnamefont {Fowler}}, \bibinfo
  {author} {\bibfnamefont {A.}~\bibnamefont {Megrant}}, \bibinfo {author}
  {\bibfnamefont {E.}~\bibnamefont {Jeffrey}}, \bibinfo {author} {\bibfnamefont
  {T.~C.}\ \bibnamefont {White}}, \bibinfo {author} {\bibfnamefont
  {D.}~\bibnamefont {Sank}}, \bibinfo {author} {\bibfnamefont {J.~Y.}\
  \bibnamefont {Mutus}}, \bibinfo {author} {\bibfnamefont {B.}~\bibnamefont
  {Campbell}}, \bibinfo {author} {\bibfnamefont {Y.}~\bibnamefont {Chen}},
  \bibinfo {author} {\bibfnamefont {Z.}~\bibnamefont {Chen}}, \bibinfo {author}
  {\bibfnamefont {B.}~\bibnamefont {Chiaro}}, \bibinfo {author} {\bibfnamefont
  {A.}~\bibnamefont {Dunsworth}}, \bibinfo {author} {\bibfnamefont {I.-C.}\
  \bibnamefont {Hoi}}, \bibinfo {author} {\bibfnamefont {C.}~\bibnamefont
  {Neill}}, \bibinfo {author} {\bibfnamefont {P.~J.~J.}\ \bibnamefont
  {O'Malley}}, \bibinfo {author} {\bibfnamefont {C.}~\bibnamefont {Quintana}},
  \bibinfo {author} {\bibfnamefont {P.}~\bibnamefont {Roushan}}, \bibinfo
  {author} {\bibfnamefont {A.}~\bibnamefont {Vainsencher}}, \bibinfo {author}
  {\bibfnamefont {J.}~\bibnamefont {Wenner}}, \bibinfo {author} {\bibfnamefont
  {A.~N.}\ \bibnamefont {Cleland}}, \ and\ \bibinfo {author} {\bibfnamefont
  {J.~M.}\ \bibnamefont {Martinis}},\ }\bibfield  {title} {\enquote {\bibinfo
  {title} {{State preservation by repetitive error detection in a
  superconducting quantum circuit}},}\ }\href {\doibase 10.1038/nature14270}
  {\bibfield  {journal} {\bibinfo  {journal} {Nature}\ }\textbf {\bibinfo
  {volume} {519}},\ \bibinfo {pages} {66} (\bibinfo {year} {2015})}\BibitemShut
  {NoStop}%
\bibitem [{\citenamefont {Takita}\ \emph {et~al.}(2017)\citenamefont {Takita},
  \citenamefont {Cross}, \citenamefont {C{\'{o}}rcoles}, \citenamefont {Chow},\
  and\ \citenamefont {Gambetta}}]{Takita2017}%
  \BibitemOpen
  \bibfield  {author} {\bibinfo {author} {\bibfnamefont {M.}~\bibnamefont
  {Takita}}, \bibinfo {author} {\bibfnamefont {A.~W.}\ \bibnamefont {Cross}},
  \bibinfo {author} {\bibfnamefont {A.~D.}\ \bibnamefont {C{\'{o}}rcoles}},
  \bibinfo {author} {\bibfnamefont {J.~M.}\ \bibnamefont {Chow}}, \ and\
  \bibinfo {author} {\bibfnamefont {J.~M.}\ \bibnamefont {Gambetta}},\
  }\bibfield  {title} {\enquote {\bibinfo {title} {{Experimental Demonstration
  of Fault-Tolerant State Preparation with Superconducting Qubits}},}\ }\href
  {\doibase 10.1103/PhysRevLett.119.180501} {\bibfield  {journal} {\bibinfo
  {journal} {Physical Review Letters}\ }\textbf {\bibinfo {volume} {119}},\
  \bibinfo {pages} {180501} (\bibinfo {year} {2017})}\BibitemShut {NoStop}%
\bibitem [{\citenamefont {Andersen}\ \emph {et~al.}(2020)\citenamefont
  {Andersen}, \citenamefont {Remm}, \citenamefont {Lazar}, \citenamefont
  {Krinner}, \citenamefont {Lacroix}, \citenamefont {Norris}, \citenamefont
  {Gabureac}, \citenamefont {Eichler},\ and\ \citenamefont
  {Wallraff}}]{CKAndersen2020}%
  \BibitemOpen
  \bibfield  {author} {\bibinfo {author} {\bibfnamefont {C.~K.}\ \bibnamefont
  {Andersen}}, \bibinfo {author} {\bibfnamefont {A.}~\bibnamefont {Remm}},
  \bibinfo {author} {\bibfnamefont {S.}~\bibnamefont {Lazar}}, \bibinfo
  {author} {\bibfnamefont {S.}~\bibnamefont {Krinner}}, \bibinfo {author}
  {\bibfnamefont {N.}~\bibnamefont {Lacroix}}, \bibinfo {author} {\bibfnamefont
  {G.~J.}\ \bibnamefont {Norris}}, \bibinfo {author} {\bibfnamefont
  {M.}~\bibnamefont {Gabureac}}, \bibinfo {author} {\bibfnamefont
  {C.}~\bibnamefont {Eichler}}, \ and\ \bibinfo {author} {\bibfnamefont
  {A.}~\bibnamefont {Wallraff}},\ }\bibfield  {title} {\enquote {\bibinfo
  {title} {Repeated quantum error detection in a surface code},}\ }\href
  {\doibase 10.1038/s41567-020-0920-y} {\bibfield  {journal} {\bibinfo
  {journal} {Nature Physics}\ }\textbf {\bibinfo {volume} {16}},\ \bibinfo
  {pages} {875} (\bibinfo {year} {2020})}\BibitemShut {NoStop}%
\bibitem [{\citenamefont {Marques}\ \emph {et~al.}(2022)\citenamefont
  {Marques}, \citenamefont {Varbanov}, \citenamefont {Moreira}, \citenamefont
  {Ali}, \citenamefont {Muthusubramanian}, \citenamefont {Zachariadis},
  \citenamefont {Battistel}, \citenamefont {Beekman}, \citenamefont {Haider},
  \citenamefont {Vlothuizen} \emph {et~al.}}]{marques2022logical}%
  \BibitemOpen
  \bibfield  {author} {\bibinfo {author} {\bibfnamefont {J.}~\bibnamefont
  {Marques}}, \bibinfo {author} {\bibfnamefont {B.}~\bibnamefont {Varbanov}},
  \bibinfo {author} {\bibfnamefont {M.}~\bibnamefont {Moreira}}, \bibinfo
  {author} {\bibfnamefont {H.}~\bibnamefont {Ali}}, \bibinfo {author}
  {\bibfnamefont {N.}~\bibnamefont {Muthusubramanian}}, \bibinfo {author}
  {\bibfnamefont {C.}~\bibnamefont {Zachariadis}}, \bibinfo {author}
  {\bibfnamefont {F.}~\bibnamefont {Battistel}}, \bibinfo {author}
  {\bibfnamefont {M.}~\bibnamefont {Beekman}}, \bibinfo {author} {\bibfnamefont
  {N.}~\bibnamefont {Haider}}, \bibinfo {author} {\bibfnamefont
  {W.}~\bibnamefont {Vlothuizen}},  \emph {et~al.},\ }\bibfield  {title}
  {\enquote {\bibinfo {title} {Logical-qubit operations in an error-detecting
  surface code},}\ }\href {\doibase 10.1038/s41567-021-01423-9} {\bibfield
  {journal} {\bibinfo  {journal} {Nature Physics}\ }\textbf {\bibinfo {volume}
  {18}},\ \bibinfo {pages} {80--86} (\bibinfo {year} {2022})}\BibitemShut
  {NoStop}%
\bibitem [{\citenamefont {Chen}\ \emph {et~al.}(2021)\citenamefont {Chen} \emph
  {et~al.}}]{Chen2021Google}%
  \BibitemOpen
  \bibfield  {author} {\bibinfo {author} {\bibfnamefont {Z.}~\bibnamefont
  {Chen}} \emph {et~al.},\ }\bibfield  {title} {\enquote {\bibinfo {title}
  {Exponential suppression of bit or phase errors with cyclic error
  correction},}\ }\href {\doibase 10.1038/s41586-021-03588-y} {\bibfield
  {journal} {\bibinfo  {journal} {Nature}\ }\textbf {\bibinfo {volume} {595}},\
  \bibinfo {pages} {383} (\bibinfo {year} {2021})}\BibitemShut {NoStop}%
\bibitem [{\citenamefont {Erhard}\ \emph {et~al.}(2021)\citenamefont {Erhard},
  \citenamefont {{Poulsen Nautrup}}, \citenamefont {Meth}, \citenamefont
  {Postler}, \citenamefont {Stricker}, \citenamefont {Stadler}, \citenamefont
  {Negnevitsky}, \citenamefont {Ringbauer}, \citenamefont {Schindler},
  \citenamefont {Briegel}, \citenamefont {Blatt}, \citenamefont {Friis},\ and\
  \citenamefont {Monz}}]{Erhard2021}%
  \BibitemOpen
  \bibfield  {author} {\bibinfo {author} {\bibfnamefont {A.}~\bibnamefont
  {Erhard}}, \bibinfo {author} {\bibfnamefont {H.}~\bibnamefont {{Poulsen
  Nautrup}}}, \bibinfo {author} {\bibfnamefont {M.}~\bibnamefont {Meth}},
  \bibinfo {author} {\bibfnamefont {L.}~\bibnamefont {Postler}}, \bibinfo
  {author} {\bibfnamefont {R.}~\bibnamefont {Stricker}}, \bibinfo {author}
  {\bibfnamefont {M.}~\bibnamefont {Stadler}}, \bibinfo {author} {\bibfnamefont
  {V.}~\bibnamefont {Negnevitsky}}, \bibinfo {author} {\bibfnamefont
  {M.}~\bibnamefont {Ringbauer}}, \bibinfo {author} {\bibfnamefont
  {P.}~\bibnamefont {Schindler}}, \bibinfo {author} {\bibfnamefont {H.~J.}\
  \bibnamefont {Briegel}}, \bibinfo {author} {\bibfnamefont {R.}~\bibnamefont
  {Blatt}}, \bibinfo {author} {\bibfnamefont {N.}~\bibnamefont {Friis}}, \ and\
  \bibinfo {author} {\bibfnamefont {T.}~\bibnamefont {Monz}},\ }\bibfield
  {title} {\enquote {\bibinfo {title} {{Entangling logical qubits with lattice
  surgery}},}\ }\href {\doibase 10.1038/s41586-020-03079-6} {\bibfield
  {journal} {\bibinfo  {journal} {Nature}\ }\textbf {\bibinfo {volume} {589}},\
  \bibinfo {pages} {220} (\bibinfo {year} {2021})}\BibitemShut {NoStop}%
\bibitem [{\citenamefont {Satzinger}\ \emph {et~al.}(2021)\citenamefont
  {Satzinger}, \citenamefont {Liu}, \citenamefont {Smith}, \citenamefont
  {Knapp}, \citenamefont {Newman}, \citenamefont {Jones}, \citenamefont {Chen},
  \citenamefont {Quintana}, \citenamefont {Mi}, \citenamefont {Dunsworth} \emph
  {et~al.}}]{satzinger2021realizing}%
  \BibitemOpen
  \bibfield  {author} {\bibinfo {author} {\bibfnamefont {K.}~\bibnamefont
  {Satzinger}}, \bibinfo {author} {\bibfnamefont {Y.-J.}\ \bibnamefont {Liu}},
  \bibinfo {author} {\bibfnamefont {A.}~\bibnamefont {Smith}}, \bibinfo
  {author} {\bibfnamefont {C.}~\bibnamefont {Knapp}}, \bibinfo {author}
  {\bibfnamefont {M.}~\bibnamefont {Newman}}, \bibinfo {author} {\bibfnamefont
  {C.}~\bibnamefont {Jones}}, \bibinfo {author} {\bibfnamefont
  {Z.}~\bibnamefont {Chen}}, \bibinfo {author} {\bibfnamefont {C.}~\bibnamefont
  {Quintana}}, \bibinfo {author} {\bibfnamefont {X.}~\bibnamefont {Mi}},
  \bibinfo {author} {\bibfnamefont {A.}~\bibnamefont {Dunsworth}},  \emph
  {et~al.},\ }\bibfield  {title} {\enquote {\bibinfo {title} {Realizing
  topologically ordered states on a quantum processor},}\ }\href {\doibase
  10.1126/science.abi8378} {\bibfield  {journal} {\bibinfo  {journal}
  {Science}\ }\textbf {\bibinfo {volume} {374}},\ \bibinfo {pages} {1237--1241}
  (\bibinfo {year} {2021})}\BibitemShut {NoStop}%
\bibitem [{\citenamefont {Egan}\ \emph {et~al.}(2021)\citenamefont {Egan},
  \citenamefont {Debroy}, \citenamefont {Noel}, \citenamefont {Risinger},
  \citenamefont {Zhu}, \citenamefont {Biswas}, \citenamefont {Newman},
  \citenamefont {Li}, \citenamefont {Brown}, \citenamefont {Cetina},\ and\
  \citenamefont {Monroe}}]{Egan2021}%
  \BibitemOpen
  \bibfield  {author} {\bibinfo {author} {\bibfnamefont {L.}~\bibnamefont
  {Egan}}, \bibinfo {author} {\bibfnamefont {D.~M.}\ \bibnamefont {Debroy}},
  \bibinfo {author} {\bibfnamefont {C.}~\bibnamefont {Noel}}, \bibinfo {author}
  {\bibfnamefont {A.}~\bibnamefont {Risinger}}, \bibinfo {author}
  {\bibfnamefont {D.}~\bibnamefont {Zhu}}, \bibinfo {author} {\bibfnamefont
  {D.}~\bibnamefont {Biswas}}, \bibinfo {author} {\bibfnamefont
  {M.}~\bibnamefont {Newman}}, \bibinfo {author} {\bibfnamefont
  {M.}~\bibnamefont {Li}}, \bibinfo {author} {\bibfnamefont {K.~R.}\
  \bibnamefont {Brown}}, \bibinfo {author} {\bibfnamefont {M.}~\bibnamefont
  {Cetina}}, \ and\ \bibinfo {author} {\bibfnamefont {C.}~\bibnamefont
  {Monroe}},\ }\bibfield  {title} {\enquote {\bibinfo {title} {{Fault-tolerant
  control of an error-corrected qubit}},}\ }\href {\doibase
  10.1038/s41586-021-03928-y} {\bibfield  {journal} {\bibinfo  {journal}
  {Nature}\ }\textbf {\bibinfo {volume} {598}},\ \bibinfo {pages} {281}
  (\bibinfo {year} {2021})}\BibitemShut {NoStop}%
\bibitem [{\citenamefont {Ryan-Anderson}\ \emph {et~al.}(2021)\citenamefont
  {Ryan-Anderson}, \citenamefont {Bohnet}, \citenamefont {Lee}, \citenamefont
  {Gresh}, \citenamefont {Hankin}, \citenamefont {Gaebler}, \citenamefont
  {Francois}, \citenamefont {Chernoguzov}, \citenamefont {Lucchetti},
  \citenamefont {Brown}, \citenamefont {Gatterman}, \citenamefont {Halit},
  \citenamefont {Gilmore}, \citenamefont {Gerber}, \citenamefont {Neyenhuis},
  \citenamefont {Hayes},\ and\ \citenamefont {Stutz}}]{Ryan-Anderson2021}%
  \BibitemOpen
  \bibfield  {author} {\bibinfo {author} {\bibfnamefont {C.}~\bibnamefont
  {Ryan-Anderson}}, \bibinfo {author} {\bibfnamefont {J.~G.}\ \bibnamefont
  {Bohnet}}, \bibinfo {author} {\bibfnamefont {K.}~\bibnamefont {Lee}},
  \bibinfo {author} {\bibfnamefont {D.}~\bibnamefont {Gresh}}, \bibinfo
  {author} {\bibfnamefont {A.}~\bibnamefont {Hankin}}, \bibinfo {author}
  {\bibfnamefont {J.~P.}\ \bibnamefont {Gaebler}}, \bibinfo {author}
  {\bibfnamefont {D.}~\bibnamefont {Francois}}, \bibinfo {author}
  {\bibfnamefont {A.}~\bibnamefont {Chernoguzov}}, \bibinfo {author}
  {\bibfnamefont {D.}~\bibnamefont {Lucchetti}}, \bibinfo {author}
  {\bibfnamefont {N.~C.}\ \bibnamefont {Brown}}, \bibinfo {author}
  {\bibfnamefont {T.~M.}\ \bibnamefont {Gatterman}}, \bibinfo {author}
  {\bibfnamefont {S.~K.}\ \bibnamefont {Halit}}, \bibinfo {author}
  {\bibfnamefont {K.}~\bibnamefont {Gilmore}}, \bibinfo {author} {\bibfnamefont
  {J.}~\bibnamefont {Gerber}}, \bibinfo {author} {\bibfnamefont
  {B.}~\bibnamefont {Neyenhuis}}, \bibinfo {author} {\bibfnamefont
  {D.}~\bibnamefont {Hayes}}, \ and\ \bibinfo {author} {\bibfnamefont {R.~P.}\
  \bibnamefont {Stutz}},\ }\href@noop {} {\enquote {\bibinfo {title}
  {{Realization of real-time fault-tolerant quantum error correction}},}\ }
  (\bibinfo {year} {2021}),\ \Eprint {http://arxiv.org/abs/2107.07505}
  {arXiv:2107.07505} \BibitemShut {NoStop}%
\bibitem [{\citenamefont {Livingston}\ \emph {et~al.}(2021)\citenamefont
  {Livingston}, \citenamefont {Blok}, \citenamefont {Flurin}, \citenamefont
  {Dressel}, \citenamefont {Jordan},\ and\ \citenamefont
  {Siddiqi}}]{Livingston2021}%
  \BibitemOpen
  \bibfield  {author} {\bibinfo {author} {\bibfnamefont {W.~P.}\ \bibnamefont
  {Livingston}}, \bibinfo {author} {\bibfnamefont {M.~S.}\ \bibnamefont
  {Blok}}, \bibinfo {author} {\bibfnamefont {E.}~\bibnamefont {Flurin}},
  \bibinfo {author} {\bibfnamefont {J.}~\bibnamefont {Dressel}}, \bibinfo
  {author} {\bibfnamefont {A.~N.}\ \bibnamefont {Jordan}}, \ and\ \bibinfo
  {author} {\bibfnamefont {I.}~\bibnamefont {Siddiqi}},\ }\bibfield  {title}
  {\enquote {\bibinfo {title} {{Experimental demonstration of continuous
  quantum error correction}},}\ }\href {http://arxiv.org/abs/2107.11398} {\
  (\bibinfo {year} {2021})},\ \Eprint {http://arxiv.org/abs/2107.11398}
  {arXiv:2107.11398} \BibitemShut {NoStop}%
\bibitem [{\citenamefont {Postler}\ \emph {et~al.}(2021)\citenamefont
  {Postler}, \citenamefont {Heußen}, \citenamefont {Pogorelov}, \citenamefont
  {Rispler}, \citenamefont {Feldker}, \citenamefont {Meth}, \citenamefont
  {Marciniak}, \citenamefont {Stricker}, \citenamefont {Ringbauer},
  \citenamefont {Blatt}, \citenamefont {Schindler}, \citenamefont {Müller},\
  and\ \citenamefont {Monz}}]{postler2021demonstration}%
  \BibitemOpen
  \bibfield  {author} {\bibinfo {author} {\bibfnamefont {L.}~\bibnamefont
  {Postler}}, \bibinfo {author} {\bibfnamefont {S.}~\bibnamefont {Heußen}},
  \bibinfo {author} {\bibfnamefont {I.}~\bibnamefont {Pogorelov}}, \bibinfo
  {author} {\bibfnamefont {M.}~\bibnamefont {Rispler}}, \bibinfo {author}
  {\bibfnamefont {T.}~\bibnamefont {Feldker}}, \bibinfo {author} {\bibfnamefont
  {M.}~\bibnamefont {Meth}}, \bibinfo {author} {\bibfnamefont {C.~D.}\
  \bibnamefont {Marciniak}}, \bibinfo {author} {\bibfnamefont {R.}~\bibnamefont
  {Stricker}}, \bibinfo {author} {\bibfnamefont {M.}~\bibnamefont {Ringbauer}},
  \bibinfo {author} {\bibfnamefont {R.}~\bibnamefont {Blatt}}, \bibinfo
  {author} {\bibfnamefont {P.}~\bibnamefont {Schindler}}, \bibinfo {author}
  {\bibfnamefont {M.}~\bibnamefont {Müller}}, \ and\ \bibinfo {author}
  {\bibfnamefont {T.}~\bibnamefont {Monz}},\ }\href@noop {} {\enquote {\bibinfo
  {title} {Demonstration of fault-tolerant universal quantum gate
  operations},}\ } (\bibinfo {year} {2021}),\ \Eprint
  {http://arxiv.org/abs/2111.12654} {arXiv:2111.12654} \BibitemShut {NoStop}%
\bibitem [{\citenamefont {Krinner}\ \emph {et~al.}(2021)\citenamefont
  {Krinner}, \citenamefont {Lacroix}, \citenamefont {Remm}, \citenamefont {{Di
  Paolo}}, \citenamefont {Genois}, \citenamefont {Leroux}, \citenamefont
  {Hellings}, \citenamefont {Lazar}, \citenamefont {Swiadek}, \citenamefont
  {Herrmann}, \citenamefont {Norris}, \citenamefont {Andersen}, \citenamefont
  {M{\"{u}}ller}, \citenamefont {Blais}, \citenamefont {Eichler},\ and\
  \citenamefont {Wallraff}}]{Krinner2021}%
  \BibitemOpen
  \bibfield  {author} {\bibinfo {author} {\bibfnamefont {S.}~\bibnamefont
  {Krinner}}, \bibinfo {author} {\bibfnamefont {N.}~\bibnamefont {Lacroix}},
  \bibinfo {author} {\bibfnamefont {A.}~\bibnamefont {Remm}}, \bibinfo {author}
  {\bibfnamefont {A.}~\bibnamefont {{Di Paolo}}}, \bibinfo {author}
  {\bibfnamefont {E.}~\bibnamefont {Genois}}, \bibinfo {author} {\bibfnamefont
  {C.}~\bibnamefont {Leroux}}, \bibinfo {author} {\bibfnamefont
  {C.}~\bibnamefont {Hellings}}, \bibinfo {author} {\bibfnamefont
  {S.}~\bibnamefont {Lazar}}, \bibinfo {author} {\bibfnamefont
  {F.}~\bibnamefont {Swiadek}}, \bibinfo {author} {\bibfnamefont
  {J.}~\bibnamefont {Herrmann}}, \bibinfo {author} {\bibfnamefont {G.~J.}\
  \bibnamefont {Norris}}, \bibinfo {author} {\bibfnamefont {C.~K.}\
  \bibnamefont {Andersen}}, \bibinfo {author} {\bibfnamefont {M.}~\bibnamefont
  {M{\"{u}}ller}}, \bibinfo {author} {\bibfnamefont {A.}~\bibnamefont {Blais}},
  \bibinfo {author} {\bibfnamefont {C.}~\bibnamefont {Eichler}}, \ and\
  \bibinfo {author} {\bibfnamefont {A.}~\bibnamefont {Wallraff}},\ }\href@noop
  {} {\enquote {\bibinfo {title} {{Realizing Repeated Quantum Error Correction
  in a Distance-Three Surface Code}},}\ } (\bibinfo {year} {2021}),\ \Eprint
  {http://arxiv.org/abs/2112.03708} {arXiv:2112.03708} \BibitemShut {NoStop}%
\bibitem [{\citenamefont {Bluvstein}\ \emph {et~al.}(2021)\citenamefont
  {Bluvstein}, \citenamefont {Levine}, \citenamefont {Semeghini}, \citenamefont
  {Wang}, \citenamefont {Ebadi}, \citenamefont {Kalinowski}, \citenamefont
  {Keesling}, \citenamefont {Maskara}, \citenamefont {Pichler}, \citenamefont
  {Greiner}, \citenamefont {Vuletic},\ and\ \citenamefont
  {Lukin}}]{Bluvstein2021}%
  \BibitemOpen
  \bibfield  {author} {\bibinfo {author} {\bibfnamefont {D.}~\bibnamefont
  {Bluvstein}}, \bibinfo {author} {\bibfnamefont {H.}~\bibnamefont {Levine}},
  \bibinfo {author} {\bibfnamefont {G.}~\bibnamefont {Semeghini}}, \bibinfo
  {author} {\bibfnamefont {T.~T.}\ \bibnamefont {Wang}}, \bibinfo {author}
  {\bibfnamefont {S.}~\bibnamefont {Ebadi}}, \bibinfo {author} {\bibfnamefont
  {M.}~\bibnamefont {Kalinowski}}, \bibinfo {author} {\bibfnamefont
  {A.}~\bibnamefont {Keesling}}, \bibinfo {author} {\bibfnamefont
  {N.}~\bibnamefont {Maskara}}, \bibinfo {author} {\bibfnamefont
  {H.}~\bibnamefont {Pichler}}, \bibinfo {author} {\bibfnamefont
  {M.}~\bibnamefont {Greiner}}, \bibinfo {author} {\bibfnamefont
  {V.}~\bibnamefont {Vuletic}}, \ and\ \bibinfo {author} {\bibfnamefont
  {M.~D.}\ \bibnamefont {Lukin}},\ }\href@noop {} {\enquote {\bibinfo {title}
  {{A quantum processor based on coherent transport of entangled atom
  arrays}},}\ } (\bibinfo {year} {2021}),\ \Eprint
  {http://arxiv.org/abs/2112.03923} {arXiv:2112.03923} \BibitemShut {NoStop}%
\bibitem [{\citenamefont {Wang}\ \emph {et~al.}(2003)\citenamefont {Wang},
  \citenamefont {Harrington},\ and\ \citenamefont
  {Preskill}}]{wang2003confinement}%
  \BibitemOpen
  \bibfield  {author} {\bibinfo {author} {\bibfnamefont {C.}~\bibnamefont
  {Wang}}, \bibinfo {author} {\bibfnamefont {J.}~\bibnamefont {Harrington}}, \
  and\ \bibinfo {author} {\bibfnamefont {J.}~\bibnamefont {Preskill}},\
  }\bibfield  {title} {\enquote {\bibinfo {title} {{Confinement-Higgs
  transition in a disordered gauge theory and the accuracy threshold for
  quantum memory}},}\ }\href {\doibase 10.1016/S0003-4916(02)00019-2}
  {\bibfield  {journal} {\bibinfo  {journal} {Annals of Physics}\ }\textbf
  {\bibinfo {volume} {303}},\ \bibinfo {pages} {31} (\bibinfo {year}
  {2003})}\BibitemShut {NoStop}%
\bibitem [{\citenamefont {Bombin}\ \emph {et~al.}(2012)\citenamefont {Bombin},
  \citenamefont {Andrist}, \citenamefont {Ohzeki}, \citenamefont {Katzgraber},\
  and\ \citenamefont {Martin-Delgado}}]{PhysRevX.2.021004}%
  \BibitemOpen
  \bibfield  {author} {\bibinfo {author} {\bibfnamefont {H.}~\bibnamefont
  {Bombin}}, \bibinfo {author} {\bibfnamefont {R.~S.}\ \bibnamefont {Andrist}},
  \bibinfo {author} {\bibfnamefont {M.}~\bibnamefont {Ohzeki}}, \bibinfo
  {author} {\bibfnamefont {H.~G.}\ \bibnamefont {Katzgraber}}, \ and\ \bibinfo
  {author} {\bibfnamefont {M.~A.}\ \bibnamefont {Martin-Delgado}},\ }\bibfield
  {title} {\enquote {\bibinfo {title} {{Strong Resilience of Topological Codes
  to Depolarization}},}\ }\href {\doibase 10.1103/PhysRevX.2.021004} {\bibfield
   {journal} {\bibinfo  {journal} {Physical Review X}\ }\textbf {\bibinfo
  {volume} {2}},\ \bibinfo {pages} {021004} (\bibinfo {year}
  {2012})}\BibitemShut {NoStop}%
\bibitem [{\citenamefont {Katzgraber}\ and\ \citenamefont
  {Andrist}(2013)}]{katzgraber2013stability}%
  \BibitemOpen
  \bibfield  {author} {\bibinfo {author} {\bibfnamefont {H.~G.}\ \bibnamefont
  {Katzgraber}}\ and\ \bibinfo {author} {\bibfnamefont {R.~S.}\ \bibnamefont
  {Andrist}},\ }\bibfield  {title} {\enquote {\bibinfo {title} {Stability of
  topologically-protected quantum computing proposals as seen through spin
  glasses},}\ }in\ \href {\doibase 10.1088/1742-6596/473/1/012019} {\emph
  {\bibinfo {booktitle} {Journal of Physics: Conference Series}}},\ Vol.\
  \bibinfo {volume} {473}\ (\bibinfo {organization} {IOP Publishing},\ \bibinfo
  {year} {2013})\ p.\ \bibinfo {pages} {012019}\BibitemShut {NoStop}%
\bibitem [{\citenamefont {Edmonds}(1965)}]{edmonds1965paths}%
  \BibitemOpen
  \bibfield  {author} {\bibinfo {author} {\bibfnamefont {J.}~\bibnamefont
  {Edmonds}},\ }\bibfield  {title} {\enquote {\bibinfo {title} {Paths, trees,
  and flowers},}\ }\href {\doibase 10.4153/CJM-1965-045-4} {\bibfield
  {journal} {\bibinfo  {journal} {Canadian Journal of Mathematics}\ }\textbf
  {\bibinfo {volume} {17}},\ \bibinfo {pages} {449} (\bibinfo {year}
  {1965})}\BibitemShut {NoStop}%
\bibitem [{\citenamefont {Wang}\ \emph {et~al.}(2010)\citenamefont {Wang},
  \citenamefont {Fowler}, \citenamefont {Stephens},\ and\ \citenamefont
  {Hollenberg}}]{wang2009threshold}%
  \BibitemOpen
  \bibfield  {author} {\bibinfo {author} {\bibfnamefont {D.~S.}\ \bibnamefont
  {Wang}}, \bibinfo {author} {\bibfnamefont {A.~G.}\ \bibnamefont {Fowler}},
  \bibinfo {author} {\bibfnamefont {A.~M.}\ \bibnamefont {Stephens}}, \ and\
  \bibinfo {author} {\bibfnamefont {L.~C.~L.}\ \bibnamefont {Hollenberg}},\
  }\bibfield  {title} {\enquote {\bibinfo {title} {{Threshold Error Rates for
  the Toric and Planar Codes}},}\ }\href {\doibase 10.5555/2011362.2011368}
  {\bibfield  {journal} {\bibinfo  {journal} {Quantum Information \&
  Computation}\ }\textbf {\bibinfo {volume} {10}},\ \bibinfo {pages} {456}
  (\bibinfo {year} {2010})}\BibitemShut {NoStop}%
\bibitem [{\citenamefont {Wang}\ \emph {et~al.}(2011)\citenamefont {Wang},
  \citenamefont {Fowler},\ and\ \citenamefont
  {Hollenberg}}]{PhysRevA.83.020302}%
  \BibitemOpen
  \bibfield  {author} {\bibinfo {author} {\bibfnamefont {D.~S.}\ \bibnamefont
  {Wang}}, \bibinfo {author} {\bibfnamefont {A.~G.}\ \bibnamefont {Fowler}}, \
  and\ \bibinfo {author} {\bibfnamefont {L.~C.~L.}\ \bibnamefont
  {Hollenberg}},\ }\bibfield  {title} {\enquote {\bibinfo {title} {Surface code
  quantum computing with error rates over 1\%},}\ }\href {\doibase
  10.1103/PhysRevA.83.020302} {\bibfield  {journal} {\bibinfo  {journal}
  {Physical Review A}\ }\textbf {\bibinfo {volume} {83}},\ \bibinfo {pages}
  {020302} (\bibinfo {year} {2011})}\BibitemShut {NoStop}%
\bibitem [{\citenamefont {Stace}\ \emph {et~al.}(2009)\citenamefont {Stace},
  \citenamefont {Barrett},\ and\ \citenamefont
  {Doherty}}]{PhysRevLett.102.200501}%
  \BibitemOpen
  \bibfield  {author} {\bibinfo {author} {\bibfnamefont {T.~M.}\ \bibnamefont
  {Stace}}, \bibinfo {author} {\bibfnamefont {S.~D.}\ \bibnamefont {Barrett}},
  \ and\ \bibinfo {author} {\bibfnamefont {A.~C.}\ \bibnamefont {Doherty}},\
  }\bibfield  {title} {\enquote {\bibinfo {title} {{Thresholds for Topological
  Codes in the Presence of Loss}},}\ }\href {\doibase
  10.1103/PhysRevLett.102.200501} {\bibfield  {journal} {\bibinfo  {journal}
  {Physical Review Letters}\ }\textbf {\bibinfo {volume} {102}},\ \bibinfo
  {pages} {200501} (\bibinfo {year} {2009})}\BibitemShut {NoStop}%
\bibitem [{\citenamefont {Stace}\ and\ \citenamefont
  {Barrett}(2010)}]{PhysRevA.81.022317}%
  \BibitemOpen
  \bibfield  {author} {\bibinfo {author} {\bibfnamefont {T.~M.}\ \bibnamefont
  {Stace}}\ and\ \bibinfo {author} {\bibfnamefont {S.~D.}\ \bibnamefont
  {Barrett}},\ }\bibfield  {title} {\enquote {\bibinfo {title} {Error
  correction and degeneracy in surface codes suffering loss},}\ }\href
  {\doibase 10.1103/PhysRevA.81.022317} {\bibfield  {journal} {\bibinfo
  {journal} {Physical Review A}\ }\textbf {\bibinfo {volume} {81}},\ \bibinfo
  {pages} {022317} (\bibinfo {year} {2010})}\BibitemShut {NoStop}%
\bibitem [{\citenamefont {Fowler}(2013{\natexlab{a}})}]{fowler2013optimal}%
  \BibitemOpen
  \bibfield  {author} {\bibinfo {author} {\bibfnamefont {A.~G.}\ \bibnamefont
  {Fowler}},\ }\href@noop {} {\enquote {\bibinfo {title} {Optimal complexity
  correction of correlated errors in the surface code},}\ } (\bibinfo {year}
  {2013}{\natexlab{a}}),\ \Eprint {http://arxiv.org/abs/1310.0863}
  {arXiv:1310.0863} \BibitemShut {NoStop}%
\bibitem [{\citenamefont {Delfosse}\ and\ \citenamefont
  {Tillich}(2014)}]{delfosse2014decoding}%
  \BibitemOpen
  \bibfield  {author} {\bibinfo {author} {\bibfnamefont {N.}~\bibnamefont
  {Delfosse}}\ and\ \bibinfo {author} {\bibfnamefont {J.-P.}\ \bibnamefont
  {Tillich}},\ }\bibfield  {title} {\enquote {\bibinfo {title} {{A decoding
  algorithm for CSS codes using the X/Z correlations}},}\ }in\ \href {\doibase
  10.1109/ISIT.2014.6874997} {\emph {\bibinfo {booktitle} {2014 IEEE
  International Symposium on Information Theory}}}\ (\bibinfo {organization}
  {IEEE},\ \bibinfo {year} {2014})\ pp.\ \bibinfo {pages}
  {1071--1075}\BibitemShut {NoStop}%
\bibitem [{\citenamefont {Fowler}(2015)}]{fowler2015minimum}%
  \BibitemOpen
  \bibfield  {author} {\bibinfo {author} {\bibfnamefont {A.~G.}\ \bibnamefont
  {Fowler}},\ }\bibfield  {title} {\enquote {\bibinfo {title} {{Minimum weight
  perfect matching of fault-tolerant topological quantum error correction in
  average O(1) parallel time}},}\ }\href
  {http://dl.acm.org/citation.cfm?id=2685188.2685197} {\bibfield  {journal}
  {\bibinfo  {journal} {Quantum Information and Computation}\ }\textbf
  {\bibinfo {volume} {15}},\ \bibinfo {pages} {145} (\bibinfo {year}
  {2015})}\BibitemShut {NoStop}%
\bibitem [{\citenamefont {Criger}\ and\ \citenamefont
  {Ashraf}(2018)}]{criger2018multi}%
  \BibitemOpen
  \bibfield  {author} {\bibinfo {author} {\bibfnamefont {B.}~\bibnamefont
  {Criger}}\ and\ \bibinfo {author} {\bibfnamefont {I.}~\bibnamefont
  {Ashraf}},\ }\bibfield  {title} {\enquote {\bibinfo {title} {{Multi-path
  summation for decoding 2D topological codes}},}\ }\href {\doibase
  10.22331/q-2018-10-19-102} {\bibfield  {journal} {\bibinfo  {journal}
  {Quantum}\ }\textbf {\bibinfo {volume} {2}},\ \bibinfo {pages} {102}
  (\bibinfo {year} {2018})}\BibitemShut {NoStop}%
\bibitem [{\citenamefont {Duclos-Cianci}\ and\ \citenamefont
  {Poulin}(2010)}]{duclos2010fast}%
  \BibitemOpen
  \bibfield  {author} {\bibinfo {author} {\bibfnamefont {G.}~\bibnamefont
  {Duclos-Cianci}}\ and\ \bibinfo {author} {\bibfnamefont {D.}~\bibnamefont
  {Poulin}},\ }\bibfield  {title} {\enquote {\bibinfo {title} {{Fast Decoders
  for Topological Quantum Codes}},}\ }\href {\doibase
  10.1103/PhysRevLett.104.050504} {\bibfield  {journal} {\bibinfo  {journal}
  {Physical Review Letters}\ }\textbf {\bibinfo {volume} {104}},\ \bibinfo
  {pages} {050504} (\bibinfo {year} {2010})}\BibitemShut {NoStop}%
\bibitem [{\citenamefont {Herold}\ \emph {et~al.}(2015)\citenamefont {Herold},
  \citenamefont {Campbell}, \citenamefont {Eisert},\ and\ \citenamefont
  {Kastoryano}}]{herold2015cellular}%
  \BibitemOpen
  \bibfield  {author} {\bibinfo {author} {\bibfnamefont {M.}~\bibnamefont
  {Herold}}, \bibinfo {author} {\bibfnamefont {E.~T.}\ \bibnamefont
  {Campbell}}, \bibinfo {author} {\bibfnamefont {J.}~\bibnamefont {Eisert}}, \
  and\ \bibinfo {author} {\bibfnamefont {M.~J.}\ \bibnamefont {Kastoryano}},\
  }\bibfield  {title} {\enquote {\bibinfo {title} {Cellular-automaton decoders
  for topological quantum memories},}\ }\href {\doibase 10.1038/npjqi.2015.10}
  {\bibfield  {journal} {\bibinfo  {journal} {npj Quantum Information}\
  }\textbf {\bibinfo {volume} {1}},\ \bibinfo {pages} {15010} (\bibinfo {year}
  {2015})}\BibitemShut {NoStop}%
\bibitem [{\citenamefont {Kubica}\ and\ \citenamefont
  {Preskill}(2019)}]{kubica2018cellular}%
  \BibitemOpen
  \bibfield  {author} {\bibinfo {author} {\bibfnamefont {A.}~\bibnamefont
  {Kubica}}\ and\ \bibinfo {author} {\bibfnamefont {J.}~\bibnamefont
  {Preskill}},\ }\bibfield  {title} {\enquote {\bibinfo {title}
  {{Cellular-Automaton Decoders with Provable Thresholds for Topological
  Codes}},}\ }\href {https://doi.org/10.1103/PhysRevLett.123.020501} {\bibfield
   {journal} {\bibinfo  {journal} {Physical Review Letters}\ }\textbf {\bibinfo
  {volume} {123}},\ \bibinfo {pages} {020501} (\bibinfo {year}
  {2019})}\BibitemShut {NoStop}%
\bibitem [{\citenamefont {Delfosse}\ and\ \citenamefont
  {Nickerson}(2021)}]{delfosse2017almost}%
  \BibitemOpen
  \bibfield  {author} {\bibinfo {author} {\bibfnamefont {N.}~\bibnamefont
  {Delfosse}}\ and\ \bibinfo {author} {\bibfnamefont {N.~H.}\ \bibnamefont
  {Nickerson}},\ }\bibfield  {title} {\enquote {\bibinfo {title} {Almost-linear
  time decoding algorithm for topological codes},}\ }\href {\doibase
  10.22331/q-2021-12-02-595} {\bibfield  {journal} {\bibinfo  {journal}
  {{Quantum}}\ }\textbf {\bibinfo {volume} {5}},\ \bibinfo {pages} {595}
  (\bibinfo {year} {2021})}\BibitemShut {NoStop}%
\bibitem [{\citenamefont {Huang}\ \emph {et~al.}(2020)\citenamefont {Huang},
  \citenamefont {Newman},\ and\ \citenamefont {Brown}}]{PhysRevA.102.012419}%
  \BibitemOpen
  \bibfield  {author} {\bibinfo {author} {\bibfnamefont {S.}~\bibnamefont
  {Huang}}, \bibinfo {author} {\bibfnamefont {M.}~\bibnamefont {Newman}}, \
  and\ \bibinfo {author} {\bibfnamefont {K.~R.}\ \bibnamefont {Brown}},\
  }\bibfield  {title} {\enquote {\bibinfo {title} {Fault-tolerant weighted
  union-find decoding on the toric code},}\ }\href {\doibase
  10.1103/PhysRevA.102.012419} {\bibfield  {journal} {\bibinfo  {journal}
  {Physical Review A}\ }\textbf {\bibinfo {volume} {102}},\ \bibinfo {pages}
  {012419} (\bibinfo {year} {2020})}\BibitemShut {NoStop}%
\bibitem [{\citenamefont {LeCun}\ \emph {et~al.}(2015)\citenamefont {LeCun},
  \citenamefont {Bengio},\ and\ \citenamefont {Hinton}}]{lecun2015deep}%
  \BibitemOpen
  \bibfield  {author} {\bibinfo {author} {\bibfnamefont {Y.}~\bibnamefont
  {LeCun}}, \bibinfo {author} {\bibfnamefont {Y.}~\bibnamefont {Bengio}}, \
  and\ \bibinfo {author} {\bibfnamefont {G.}~\bibnamefont {Hinton}},\
  }\bibfield  {title} {\enquote {\bibinfo {title} {Deep learning},}\ }\href
  {\doibase 10.1038/nature14539} {\bibfield  {journal} {\bibinfo  {journal}
  {Nature}\ }\textbf {\bibinfo {volume} {521}},\ \bibinfo {pages} {436}
  (\bibinfo {year} {2015})}\BibitemShut {NoStop}%
\bibitem [{\citenamefont {Goodfellow}\ \emph {et~al.}(2016)\citenamefont
  {Goodfellow}, \citenamefont {Bengio}, \citenamefont {Courville},\ and\
  \citenamefont {Bengio}}]{goodfellow2016deep}%
  \BibitemOpen
  \bibfield  {author} {\bibinfo {author} {\bibfnamefont {I.}~\bibnamefont
  {Goodfellow}}, \bibinfo {author} {\bibfnamefont {Y.}~\bibnamefont {Bengio}},
  \bibinfo {author} {\bibfnamefont {A.}~\bibnamefont {Courville}}, \ and\
  \bibinfo {author} {\bibfnamefont {Y.}~\bibnamefont {Bengio}},\ }\href@noop {}
  {\emph {\bibinfo {title} {Deep learning}}}\ (\bibinfo  {publisher} {MIT press
  Cambridge},\ \bibinfo {year} {2016})\BibitemShut {NoStop}%
\bibitem [{\citenamefont {Carleo}\ and\ \citenamefont
  {Troyer}(2017)}]{carleo2017solving}%
  \BibitemOpen
  \bibfield  {author} {\bibinfo {author} {\bibfnamefont {G.}~\bibnamefont
  {Carleo}}\ and\ \bibinfo {author} {\bibfnamefont {M.}~\bibnamefont
  {Troyer}},\ }\bibfield  {title} {\enquote {\bibinfo {title} {Solving the
  quantum many-body problem with artificial neural networks},}\ }\href
  {\doibase 10.1126/science.aag2302} {\bibfield  {journal} {\bibinfo  {journal}
  {Science}\ }\textbf {\bibinfo {volume} {355}},\ \bibinfo {pages} {602}
  (\bibinfo {year} {2017})}\BibitemShut {NoStop}%
\bibitem [{\citenamefont {Carrasquilla}\ and\ \citenamefont
  {Melko}(2017)}]{carrasquilla2017machine}%
  \BibitemOpen
  \bibfield  {author} {\bibinfo {author} {\bibfnamefont {J.}~\bibnamefont
  {Carrasquilla}}\ and\ \bibinfo {author} {\bibfnamefont {R.~G.}\ \bibnamefont
  {Melko}},\ }\bibfield  {title} {\enquote {\bibinfo {title} {Machine learning
  phases of matter},}\ }\href {\doibase 10.1038/nphys4035} {\bibfield
  {journal} {\bibinfo  {journal} {Nature Physics}\ }\textbf {\bibinfo {volume}
  {13}},\ \bibinfo {pages} {431} (\bibinfo {year} {2017})}\BibitemShut
  {NoStop}%
\bibitem [{\citenamefont {Van~Nieuwenburg}\ \emph {et~al.}(2017)\citenamefont
  {Van~Nieuwenburg}, \citenamefont {Liu},\ and\ \citenamefont
  {Huber}}]{van2017learning}%
  \BibitemOpen
  \bibfield  {author} {\bibinfo {author} {\bibfnamefont {E.~P.}\ \bibnamefont
  {Van~Nieuwenburg}}, \bibinfo {author} {\bibfnamefont {Y.-H.}\ \bibnamefont
  {Liu}}, \ and\ \bibinfo {author} {\bibfnamefont {S.~D.}\ \bibnamefont
  {Huber}},\ }\bibfield  {title} {\enquote {\bibinfo {title} {Learning phase
  transitions by confusion},}\ }\href {\doibase 10.1038/nphys4037} {\bibfield
  {journal} {\bibinfo  {journal} {Nature Physics}\ }\textbf {\bibinfo {volume}
  {13}},\ \bibinfo {pages} {435} (\bibinfo {year} {2017})}\BibitemShut
  {NoStop}%
\bibitem [{\citenamefont {Carrasquilla}(2020)}]{carrasquilla2020machine}%
  \BibitemOpen
  \bibfield  {author} {\bibinfo {author} {\bibfnamefont {J.}~\bibnamefont
  {Carrasquilla}},\ }\bibfield  {title} {\enquote {\bibinfo {title} {Machine
  learning for quantum matter},}\ }\href {\doibase
  10.1080/23746149.2020.1797528} {\bibfield  {journal} {\bibinfo  {journal}
  {Advances in Physics: X}\ }\textbf {\bibinfo {volume} {5}},\ \bibinfo {pages}
  {1797528} (\bibinfo {year} {2020})}\BibitemShut {NoStop}%
\bibitem [{\citenamefont {Ahmed}\ \emph {et~al.}(2021)\citenamefont {Ahmed},
  \citenamefont {{S{\'{a}}nchez Mu{\~{n}}oz}}, \citenamefont {Nori},\ and\
  \citenamefont {Kockum}}]{Ahmed2021}%
  \BibitemOpen
  \bibfield  {author} {\bibinfo {author} {\bibfnamefont {S.}~\bibnamefont
  {Ahmed}}, \bibinfo {author} {\bibfnamefont {C.}~\bibnamefont {{S{\'{a}}nchez
  Mu{\~{n}}oz}}}, \bibinfo {author} {\bibfnamefont {F.}~\bibnamefont {Nori}}, \
  and\ \bibinfo {author} {\bibfnamefont {A.~F.}\ \bibnamefont {Kockum}},\
  }\bibfield  {title} {\enquote {\bibinfo {title} {{Quantum State Tomography
  with Conditional Generative Adversarial Networks}},}\ }\href {\doibase
  10.1103/PhysRevLett.127.140502} {\bibfield  {journal} {\bibinfo  {journal}
  {Physical Review Letters}\ }\textbf {\bibinfo {volume} {127}},\ \bibinfo
  {pages} {140502} (\bibinfo {year} {2021})}\BibitemShut {NoStop}%
\bibitem [{\citenamefont {Sweke}\ \emph {et~al.}(2020)\citenamefont {Sweke},
  \citenamefont {Kesselring}, \citenamefont {van Nieuwenburg},\ and\
  \citenamefont {Eisert}}]{sweke2020reinforcement}%
  \BibitemOpen
  \bibfield  {author} {\bibinfo {author} {\bibfnamefont {R.}~\bibnamefont
  {Sweke}}, \bibinfo {author} {\bibfnamefont {M.~S.}\ \bibnamefont
  {Kesselring}}, \bibinfo {author} {\bibfnamefont {E.~P.}\ \bibnamefont {van
  Nieuwenburg}}, \ and\ \bibinfo {author} {\bibfnamefont {J.}~\bibnamefont
  {Eisert}},\ }\bibfield  {title} {\enquote {\bibinfo {title} {Reinforcement
  learning decoders for fault-tolerant quantum computation},}\ }\href {\doibase
  10.1088/2632-2153/abc609} {\bibfield  {journal} {\bibinfo  {journal} {Machine
  Learning: Science and Technology}\ }\textbf {\bibinfo {volume} {2}},\
  \bibinfo {pages} {025005} (\bibinfo {year} {2020})}\BibitemShut {NoStop}%
\bibitem [{\citenamefont {Andreasson}\ \emph {et~al.}(2019)\citenamefont
  {Andreasson}, \citenamefont {Johansson}, \citenamefont {Liljestrand},\ and\
  \citenamefont {Granath}}]{andreasson2018quantum}%
  \BibitemOpen
  \bibfield  {author} {\bibinfo {author} {\bibfnamefont {P.}~\bibnamefont
  {Andreasson}}, \bibinfo {author} {\bibfnamefont {J.}~\bibnamefont
  {Johansson}}, \bibinfo {author} {\bibfnamefont {S.}~\bibnamefont
  {Liljestrand}}, \ and\ \bibinfo {author} {\bibfnamefont {M.}~\bibnamefont
  {Granath}},\ }\bibfield  {title} {\enquote {\bibinfo {title} {Quantum error
  correction for the toric code using deep reinforcement learning},}\ }\href
  {\doibase 10.22331/q-2019-09-02-183} {\bibfield  {journal} {\bibinfo
  {journal} {Quantum}\ }\textbf {\bibinfo {volume} {3}},\ \bibinfo {pages}
  {183} (\bibinfo {year} {2019})}\BibitemShut {NoStop}%
\bibitem [{\citenamefont {Nautrup}\ \emph {et~al.}(2019)\citenamefont
  {Nautrup}, \citenamefont {Delfosse}, \citenamefont {Dunjko}, \citenamefont
  {Briegel},\ and\ \citenamefont {Friis}}]{nautrup2018optimizing}%
  \BibitemOpen
  \bibfield  {author} {\bibinfo {author} {\bibfnamefont {H.~P.}\ \bibnamefont
  {Nautrup}}, \bibinfo {author} {\bibfnamefont {N.}~\bibnamefont {Delfosse}},
  \bibinfo {author} {\bibfnamefont {V.}~\bibnamefont {Dunjko}}, \bibinfo
  {author} {\bibfnamefont {H.~J.}\ \bibnamefont {Briegel}}, \ and\ \bibinfo
  {author} {\bibfnamefont {N.}~\bibnamefont {Friis}},\ }\bibfield  {title}
  {\enquote {\bibinfo {title} {{Optimizing Quantum Error Correction Codes with
  Reinforcement Learning}},}\ }\href {\doibase 10.22331/q-2019-12-16-215}
  {\bibfield  {journal} {\bibinfo  {journal} {Quantum}\ }\textbf {\bibinfo
  {volume} {3}},\ \bibinfo {pages} {215} (\bibinfo {year} {2019})}\BibitemShut
  {NoStop}%
\bibitem [{\citenamefont {Colomer}\ \emph {et~al.}(2020)\citenamefont
  {Colomer}, \citenamefont {Skotiniotis},\ and\ \citenamefont
  {Mu{\~n}oz-Tapia}}]{colomer2020reinforcement}%
  \BibitemOpen
  \bibfield  {author} {\bibinfo {author} {\bibfnamefont {L.~D.}\ \bibnamefont
  {Colomer}}, \bibinfo {author} {\bibfnamefont {M.}~\bibnamefont
  {Skotiniotis}}, \ and\ \bibinfo {author} {\bibfnamefont {R.}~\bibnamefont
  {Mu{\~n}oz-Tapia}},\ }\bibfield  {title} {\enquote {\bibinfo {title}
  {Reinforcement learning for optimal error correction of toric codes},}\
  }\href {\doibase 10.1016/j.physleta.2020.126353} {\bibfield  {journal}
  {\bibinfo  {journal} {Physics Letters A}\ }\textbf {\bibinfo {volume}
  {384}},\ \bibinfo {pages} {126353} (\bibinfo {year} {2020})}\BibitemShut
  {NoStop}%
\bibitem [{\citenamefont {Fitzek}\ \emph {et~al.}(2020)\citenamefont {Fitzek},
  \citenamefont {Eliasson}, \citenamefont {Kockum},\ and\ \citenamefont
  {Granath}}]{Fitzek_DRL}%
  \BibitemOpen
  \bibfield  {author} {\bibinfo {author} {\bibfnamefont {D.}~\bibnamefont
  {Fitzek}}, \bibinfo {author} {\bibfnamefont {M.}~\bibnamefont {Eliasson}},
  \bibinfo {author} {\bibfnamefont {A.~F.}\ \bibnamefont {Kockum}}, \ and\
  \bibinfo {author} {\bibfnamefont {M.}~\bibnamefont {Granath}},\ }\bibfield
  {title} {\enquote {\bibinfo {title} {{Deep Q-learning decoder for
  depolarizing noise on the toric code}},}\ }\href {\doibase
  10.1103/PhysRevResearch.2.023230} {\bibfield  {journal} {\bibinfo  {journal}
  {Physical Review Research}\ }\textbf {\bibinfo {volume} {2}},\ \bibinfo
  {pages} {023230} (\bibinfo {year} {2020})}\BibitemShut {NoStop}%
\bibitem [{\citenamefont {Théveniaut}\ and\ \citenamefont {van
  Nieuwenburg}(2021)}]{10.21468/SciPostPhys.11.1.005}%
  \BibitemOpen
  \bibfield  {author} {\bibinfo {author} {\bibfnamefont {H.}~\bibnamefont
  {Théveniaut}}\ and\ \bibinfo {author} {\bibfnamefont {E.}~\bibnamefont {van
  Nieuwenburg}},\ }\bibfield  {title} {\enquote {\bibinfo {title} {{A NEAT
  Quantum Error Decoder}},}\ }\href {\doibase 10.21468/SciPostPhys.11.1.005}
  {\bibfield  {journal} {\bibinfo  {journal} {SciPost Physics}\ }\textbf
  {\bibinfo {volume} {11}},\ \bibinfo {pages} {5} (\bibinfo {year}
  {2021})}\BibitemShut {NoStop}%
\bibitem [{\citenamefont {Torlai}\ and\ \citenamefont
  {Melko}(2017)}]{torlai2017neural}%
  \BibitemOpen
  \bibfield  {author} {\bibinfo {author} {\bibfnamefont {G.}~\bibnamefont
  {Torlai}}\ and\ \bibinfo {author} {\bibfnamefont {R.~G.}\ \bibnamefont
  {Melko}},\ }\bibfield  {title} {\enquote {\bibinfo {title} {{Neural Decoder
  for Topological Codes}},}\ }\href {\doibase 10.1103/PhysRevLett.119.030501}
  {\bibfield  {journal} {\bibinfo  {journal} {Physical Review Letters}\
  }\textbf {\bibinfo {volume} {119}},\ \bibinfo {pages} {030501} (\bibinfo
  {year} {2017})}\BibitemShut {NoStop}%
\bibitem [{\citenamefont {Krastanov}\ and\ \citenamefont
  {Jiang}(2017)}]{krastanov2017deep}%
  \BibitemOpen
  \bibfield  {author} {\bibinfo {author} {\bibfnamefont {S.}~\bibnamefont
  {Krastanov}}\ and\ \bibinfo {author} {\bibfnamefont {L.}~\bibnamefont
  {Jiang}},\ }\bibfield  {title} {\enquote {\bibinfo {title} {Deep neural
  network probabilistic decoder for stabilizer codes},}\ }\href {\doibase
  10.1038/s41598-017-11266-1} {\bibfield  {journal} {\bibinfo  {journal}
  {Scientific Reports}\ }\textbf {\bibinfo {volume} {7}},\ \bibinfo {pages}
  {11003} (\bibinfo {year} {2017})}\BibitemShut {NoStop}%
\bibitem [{\citenamefont {Varsamopoulos}\ \emph {et~al.}(2017)\citenamefont
  {Varsamopoulos}, \citenamefont {Criger},\ and\ \citenamefont
  {Bertels}}]{varsamopoulos2017decoding}%
  \BibitemOpen
  \bibfield  {author} {\bibinfo {author} {\bibfnamefont {S.}~\bibnamefont
  {Varsamopoulos}}, \bibinfo {author} {\bibfnamefont {B.}~\bibnamefont
  {Criger}}, \ and\ \bibinfo {author} {\bibfnamefont {K.}~\bibnamefont
  {Bertels}},\ }\bibfield  {title} {\enquote {\bibinfo {title} {Decoding small
  surface codes with feedforward neural networks},}\ }\href {\doibase
  10.1088/2058-9565/aa955a} {\bibfield  {journal} {\bibinfo  {journal} {Quantum
  Science and Technology}\ }\textbf {\bibinfo {volume} {3}},\ \bibinfo {pages}
  {015004} (\bibinfo {year} {2017})}\BibitemShut {NoStop}%
\bibitem [{\citenamefont {Baireuther}\ \emph {et~al.}(2018)\citenamefont
  {Baireuther}, \citenamefont {O'Brien}, \citenamefont {Tarasinski},\ and\
  \citenamefont {Beenakker}}]{baireuther2018machine}%
  \BibitemOpen
  \bibfield  {author} {\bibinfo {author} {\bibfnamefont {P.}~\bibnamefont
  {Baireuther}}, \bibinfo {author} {\bibfnamefont {T.~E.}\ \bibnamefont
  {O'Brien}}, \bibinfo {author} {\bibfnamefont {B.}~\bibnamefont {Tarasinski}},
  \ and\ \bibinfo {author} {\bibfnamefont {C.~W.}\ \bibnamefont {Beenakker}},\
  }\bibfield  {title} {\enquote {\bibinfo {title} {Machine-learning-assisted
  correction of correlated qubit errors in a topological code},}\ }\href
  {\doibase 10.22331/q-2018-01-29-48} {\bibfield  {journal} {\bibinfo
  {journal} {Quantum}\ }\textbf {\bibinfo {volume} {2}},\ \bibinfo {pages} {48}
  (\bibinfo {year} {2018})}\BibitemShut {NoStop}%
\bibitem [{\citenamefont {Breuckmann}\ and\ \citenamefont
  {Ni}(2018)}]{breuckmann2018scalable}%
  \BibitemOpen
  \bibfield  {author} {\bibinfo {author} {\bibfnamefont {N.~P.}\ \bibnamefont
  {Breuckmann}}\ and\ \bibinfo {author} {\bibfnamefont {X.}~\bibnamefont
  {Ni}},\ }\bibfield  {title} {\enquote {\bibinfo {title} {{Scalable Neural
  Network Decoders for Higher Dimensional Quantum Codes}},}\ }\href {\doibase
  10.22331/q-2018-05-24-68} {\bibfield  {journal} {\bibinfo  {journal}
  {Quantum}\ }\textbf {\bibinfo {volume} {2}},\ \bibinfo {pages} {68} (\bibinfo
  {year} {2018})}\BibitemShut {NoStop}%
\bibitem [{\citenamefont {Chamberland}\ and\ \citenamefont
  {Ronagh}(2018)}]{chamberland2018deep}%
  \BibitemOpen
  \bibfield  {author} {\bibinfo {author} {\bibfnamefont {C.}~\bibnamefont
  {Chamberland}}\ and\ \bibinfo {author} {\bibfnamefont {P.}~\bibnamefont
  {Ronagh}},\ }\bibfield  {title} {\enquote {\bibinfo {title} {Deep neural
  decoders for near term fault-tolerant experiments},}\ }\href {\doibase
  10.1088/2058-9565/aad1f7} {\bibfield  {journal} {\bibinfo  {journal} {Quantum
  Science and Technology}\ }\textbf {\bibinfo {volume} {3}},\ \bibinfo {pages}
  {044002} (\bibinfo {year} {2018})}\BibitemShut {NoStop}%
\bibitem [{\citenamefont {Ni}(2020)}]{Ni2020neuralnetwork}%
  \BibitemOpen
  \bibfield  {author} {\bibinfo {author} {\bibfnamefont {X.}~\bibnamefont
  {Ni}},\ }\bibfield  {title} {\enquote {\bibinfo {title} {Neural {N}etwork
  {D}ecoders for {L}arge-{D}istance 2{D} {T}oric {C}odes},}\ }\href {\doibase
  10.22331/q-2020-08-24-310} {\bibfield  {journal} {\bibinfo  {journal}
  {{Quantum}}\ }\textbf {\bibinfo {volume} {4}},\ \bibinfo {pages} {310}
  (\bibinfo {year} {2020})}\BibitemShut {NoStop}%
\bibitem [{\citenamefont {Gicev}\ \emph {et~al.}(2021)\citenamefont {Gicev},
  \citenamefont {Hollenberg},\ and\ \citenamefont {Usman}}]{gicev2021scalable}%
  \BibitemOpen
  \bibfield  {author} {\bibinfo {author} {\bibfnamefont {S.}~\bibnamefont
  {Gicev}}, \bibinfo {author} {\bibfnamefont {L.~C.}\ \bibnamefont
  {Hollenberg}}, \ and\ \bibinfo {author} {\bibfnamefont {M.}~\bibnamefont
  {Usman}},\ }\href@noop {} {\enquote {\bibinfo {title} {A scalable and fast
  artificial neural network syndrome decoder for surface codes},}\ } (\bibinfo
  {year} {2021}),\ \Eprint {http://arxiv.org/abs/2110.05854} {arXiv:2110.05854}
  \BibitemShut {NoStop}%
\bibitem [{\citenamefont {Wootton}\ and\ \citenamefont
  {Loss}(2012)}]{wootton2012}%
  \BibitemOpen
  \bibfield  {author} {\bibinfo {author} {\bibfnamefont {J.~R.}\ \bibnamefont
  {Wootton}}\ and\ \bibinfo {author} {\bibfnamefont {D.}~\bibnamefont {Loss}},\
  }\bibfield  {title} {\enquote {\bibinfo {title} {{High Threshold Error
  Correction for the Surface Code}},}\ }\href {\doibase
  10.1103/PhysRevLett.109.160503} {\bibfield  {journal} {\bibinfo  {journal}
  {Physical Review Letters}\ }\textbf {\bibinfo {volume} {109}},\ \bibinfo
  {pages} {160503} (\bibinfo {year} {2012})}\BibitemShut {NoStop}%
\bibitem [{\citenamefont {Hutter}\ \emph {et~al.}(2014)\citenamefont {Hutter},
  \citenamefont {Wootton},\ and\ \citenamefont {Loss}}]{hutter2014efficient}%
  \BibitemOpen
  \bibfield  {author} {\bibinfo {author} {\bibfnamefont {A.}~\bibnamefont
  {Hutter}}, \bibinfo {author} {\bibfnamefont {J.~R.}\ \bibnamefont {Wootton}},
  \ and\ \bibinfo {author} {\bibfnamefont {D.}~\bibnamefont {Loss}},\
  }\bibfield  {title} {\enquote {\bibinfo {title} {{Efficient Markov chain
  Monte Carlo algorithm for the surface code}},}\ }\href {\doibase
  10.1103/PhysRevA.89.022326} {\bibfield  {journal} {\bibinfo  {journal}
  {Physical Review A}\ }\textbf {\bibinfo {volume} {89}},\ \bibinfo {pages}
  {022326} (\bibinfo {year} {2014})}\BibitemShut {NoStop}%
\bibitem [{\citenamefont {Bravyi}\ \emph {et~al.}(2014)\citenamefont {Bravyi},
  \citenamefont {Suchara},\ and\ \citenamefont {Vargo}}]{Bravyi2014}%
  \BibitemOpen
  \bibfield  {author} {\bibinfo {author} {\bibfnamefont {S.}~\bibnamefont
  {Bravyi}}, \bibinfo {author} {\bibfnamefont {M.}~\bibnamefont {Suchara}}, \
  and\ \bibinfo {author} {\bibfnamefont {A.}~\bibnamefont {Vargo}},\ }\bibfield
   {title} {\enquote {\bibinfo {title} {Efficient algorithms for maximum
  likelihood decoding in the surface code},}\ }\href {\doibase
  10.1103/PhysRevA.90.032326} {\bibfield  {journal} {\bibinfo  {journal}
  {Physical Review A}\ }\textbf {\bibinfo {volume} {90}},\ \bibinfo {pages}
  {032326} (\bibinfo {year} {2014})}\BibitemShut {NoStop}%
\bibitem [{\citenamefont {Bombin}\ and\ \citenamefont
  {Martin-Delgado}(2007)}]{PhysRevA.76.012305}%
  \BibitemOpen
  \bibfield  {author} {\bibinfo {author} {\bibfnamefont {H.}~\bibnamefont
  {Bombin}}\ and\ \bibinfo {author} {\bibfnamefont {M.~A.}\ \bibnamefont
  {Martin-Delgado}},\ }\bibfield  {title} {\enquote {\bibinfo {title} {{Optimal
  resources for topological two-dimensional stabilizer codes: Comparative
  study}},}\ }\href {\doibase 10.1103/PhysRevA.76.012305} {\bibfield  {journal}
  {\bibinfo  {journal} {Physical Review A}\ }\textbf {\bibinfo {volume} {76}},\
  \bibinfo {pages} {012305} (\bibinfo {year} {2007})}\BibitemShut {NoStop}%
\bibitem [{\citenamefont {Tuckett}\ \emph {et~al.}(2019)\citenamefont
  {Tuckett}, \citenamefont {Darmawan}, \citenamefont {Chubb}, \citenamefont
  {Bravyi}, \citenamefont {Bartlett},\ and\ \citenamefont
  {Flammia}}]{PhysRevX.9.041031}%
  \BibitemOpen
  \bibfield  {author} {\bibinfo {author} {\bibfnamefont {D.~K.}\ \bibnamefont
  {Tuckett}}, \bibinfo {author} {\bibfnamefont {A.~S.}\ \bibnamefont
  {Darmawan}}, \bibinfo {author} {\bibfnamefont {C.~T.}\ \bibnamefont {Chubb}},
  \bibinfo {author} {\bibfnamefont {S.}~\bibnamefont {Bravyi}}, \bibinfo
  {author} {\bibfnamefont {S.~D.}\ \bibnamefont {Bartlett}}, \ and\ \bibinfo
  {author} {\bibfnamefont {S.~T.}\ \bibnamefont {Flammia}},\ }\bibfield
  {title} {\enquote {\bibinfo {title} {{Tailoring Surface Codes for Highly
  Biased Noise}},}\ }\href {\doibase 10.1103/PhysRevX.9.041031} {\bibfield
  {journal} {\bibinfo  {journal} {Physical Review X}\ }\textbf {\bibinfo
  {volume} {9}},\ \bibinfo {pages} {041031} (\bibinfo {year}
  {2019})}\BibitemShut {NoStop}%
\bibitem [{\citenamefont {Ataides}\ \emph {et~al.}(2021)\citenamefont
  {Ataides}, \citenamefont {Tuckett}, \citenamefont {Bartlett}, \citenamefont
  {Flammia},\ and\ \citenamefont {Brown}}]{bonilla2020xzzx}%
  \BibitemOpen
  \bibfield  {author} {\bibinfo {author} {\bibfnamefont {J.~P.~B.}\
  \bibnamefont {Ataides}}, \bibinfo {author} {\bibfnamefont {D.~K.}\
  \bibnamefont {Tuckett}}, \bibinfo {author} {\bibfnamefont {S.~D.}\
  \bibnamefont {Bartlett}}, \bibinfo {author} {\bibfnamefont {S.~T.}\
  \bibnamefont {Flammia}}, \ and\ \bibinfo {author} {\bibfnamefont {B.~J.}\
  \bibnamefont {Brown}},\ }\bibfield  {title} {\enquote {\bibinfo {title} {{The
  XZZX surface code}},}\ }\href {\doibase 10.1038/s41467-021-22274-1}
  {\bibfield  {journal} {\bibinfo  {journal} {Nature Communications}\ }\textbf
  {\bibinfo {volume} {12}},\ \bibinfo {pages} {2172} (\bibinfo {year}
  {2021})}\BibitemShut {NoStop}%
\bibitem [{\citenamefont {Wen}(2003)}]{PhysRevLett.90.016803}%
  \BibitemOpen
  \bibfield  {author} {\bibinfo {author} {\bibfnamefont {X.-G.}\ \bibnamefont
  {Wen}},\ }\bibfield  {title} {\enquote {\bibinfo {title} {{Quantum Orders in
  an Exact Soluble Model}},}\ }\href {\doibase 10.1103/PhysRevLett.90.016803}
  {\bibfield  {journal} {\bibinfo  {journal} {Physical Review Letters}\
  }\textbf {\bibinfo {volume} {90}},\ \bibinfo {pages} {016803} (\bibinfo
  {year} {2003})}\BibitemShut {NoStop}%
\bibitem [{\citenamefont {Kay}(2011)}]{PhysRevLett.107.270502}%
  \BibitemOpen
  \bibfield  {author} {\bibinfo {author} {\bibfnamefont {A.}~\bibnamefont
  {Kay}},\ }\bibfield  {title} {\enquote {\bibinfo {title} {{Capabilities of a
  Perturbed Toric Code as a Quantum Memory}},}\ }\href {\doibase
  10.1103/PhysRevLett.107.270502} {\bibfield  {journal} {\bibinfo  {journal}
  {Physical Review Letters}\ }\textbf {\bibinfo {volume} {107}},\ \bibinfo
  {pages} {270502} (\bibinfo {year} {2011})}\BibitemShut {NoStop}%
\bibitem [{\citenamefont {Geller}\ and\ \citenamefont
  {Zhou}(2013)}]{PhysRevA.88.012314}%
  \BibitemOpen
  \bibfield  {author} {\bibinfo {author} {\bibfnamefont {M.~R.}\ \bibnamefont
  {Geller}}\ and\ \bibinfo {author} {\bibfnamefont {Z.}~\bibnamefont {Zhou}},\
  }\bibfield  {title} {\enquote {\bibinfo {title} {{Efficient error models for
  fault-tolerant architectures and the Pauli twirling approximation}},}\ }\href
  {\doibase 10.1103/PhysRevA.88.012314} {\bibfield  {journal} {\bibinfo
  {journal} {Physical Review A}\ }\textbf {\bibinfo {volume} {88}},\ \bibinfo
  {pages} {012314} (\bibinfo {year} {2013})}\BibitemShut {NoStop}%
\bibitem [{\citenamefont {Tomita}\ and\ \citenamefont
  {Svore}(2014)}]{Tomita2014}%
  \BibitemOpen
  \bibfield  {author} {\bibinfo {author} {\bibfnamefont {Y.}~\bibnamefont
  {Tomita}}\ and\ \bibinfo {author} {\bibfnamefont {K.~M.}\ \bibnamefont
  {Svore}},\ }\bibfield  {title} {\enquote {\bibinfo {title} {Low-distance
  surface codes under realistic quantum noise},}\ }\href {\doibase
  10.1103/PhysRevA.90.062320} {\bibfield  {journal} {\bibinfo  {journal}
  {Physical Review A}\ }\textbf {\bibinfo {volume} {90}},\ \bibinfo {pages}
  {062320} (\bibinfo {year} {2014})}\BibitemShut {NoStop}%
\bibitem [{\citenamefont {Stephens}(2014)}]{PhysRevA.89.022321}%
  \BibitemOpen
  \bibfield  {author} {\bibinfo {author} {\bibfnamefont {A.~M.}\ \bibnamefont
  {Stephens}},\ }\bibfield  {title} {\enquote {\bibinfo {title} {Fault-tolerant
  thresholds for quantum error correction with the surface code},}\ }\href
  {\doibase 10.1103/PhysRevA.89.022321} {\bibfield  {journal} {\bibinfo
  {journal} {Physical Review A}\ }\textbf {\bibinfo {volume} {89}},\ \bibinfo
  {pages} {022321} (\bibinfo {year} {2014})}\BibitemShut {NoStop}%
\bibitem [{\citenamefont {Fowler}(2013{\natexlab{b}})}]{PhysRevA.88.042308}%
  \BibitemOpen
  \bibfield  {author} {\bibinfo {author} {\bibfnamefont {A.~G.}\ \bibnamefont
  {Fowler}},\ }\bibfield  {title} {\enquote {\bibinfo {title} {Coping with
  qubit leakage in topological codes},}\ }\href {\doibase
  10.1103/PhysRevA.88.042308} {\bibfield  {journal} {\bibinfo  {journal}
  {Physical Review A}\ }\textbf {\bibinfo {volume} {88}},\ \bibinfo {pages}
  {042308} (\bibinfo {year} {2013}{\natexlab{b}})}\BibitemShut {NoStop}%
\bibitem [{\citenamefont {O’Brien}\ \emph {et~al.}(2017)\citenamefont
  {O’Brien}, \citenamefont {Tarasinski},\ and\ \citenamefont
  {DiCarlo}}]{o2017density}%
  \BibitemOpen
  \bibfield  {author} {\bibinfo {author} {\bibfnamefont {T.}~\bibnamefont
  {O’Brien}}, \bibinfo {author} {\bibfnamefont {B.}~\bibnamefont
  {Tarasinski}}, \ and\ \bibinfo {author} {\bibfnamefont {L.}~\bibnamefont
  {DiCarlo}},\ }\bibfield  {title} {\enquote {\bibinfo {title} {Density-matrix
  simulation of small surface codes under current and projected experimental
  noise},}\ }\href {\doibase 10.1038/s41534-017-0039-x} {\bibfield  {journal}
  {\bibinfo  {journal} {npj Quantum Information}\ }\textbf {\bibinfo {volume}
  {3}},\ \bibinfo {pages} {39} (\bibinfo {year} {2017})}\BibitemShut {NoStop}%
\bibitem [{Note1()}]{Note1}%
  \BibitemOpen
  \bibinfo {note} {In order to uniquely specify the equivalence class, it is
  necessary to choose a specific representation of the logical operators. As
  the error chain does not commute with all stabilizers, deforming a logical
  operator may change the equivalence class classification. For the surface
  code we choose $X_L$ as the product of $X$ operators on the left edge and
  $Z_L$ as the product of $Z$ operators on the top edge, such that the class of
  the chain can be identified from the the parity of $Z$ errors on the left
  edge and the parity of $X$ errors on the top edge. Similarly for the XZZX
  code, where the operators on the edges are mixed.}\BibitemShut {Stop}%
\bibitem [{\citenamefont {Tuckett}\ \emph {et~al.}(2018)\citenamefont
  {Tuckett}, \citenamefont {Bartlett},\ and\ \citenamefont
  {Flammia}}]{PhysRevLett.120.050505}%
  \BibitemOpen
  \bibfield  {author} {\bibinfo {author} {\bibfnamefont {D.~K.}\ \bibnamefont
  {Tuckett}}, \bibinfo {author} {\bibfnamefont {S.~D.}\ \bibnamefont
  {Bartlett}}, \ and\ \bibinfo {author} {\bibfnamefont {S.~T.}\ \bibnamefont
  {Flammia}},\ }\bibfield  {title} {\enquote {\bibinfo {title} {{Ultrahigh
  Error Threshold for Surface Codes with Biased Noise}},}\ }\href {\doibase
  10.1103/PhysRevLett.120.050505} {\bibfield  {journal} {\bibinfo  {journal}
  {Physical Review Letters}\ }\textbf {\bibinfo {volume} {120}},\ \bibinfo
  {pages} {050505} (\bibinfo {year} {2018})}\BibitemShut {NoStop}%
\bibitem [{\citenamefont {Tuckett}(2020)}]{qecsim}%
  \BibitemOpen
  \bibfield  {author} {\bibinfo {author} {\bibfnamefont {D.~K.}\ \bibnamefont
  {Tuckett}},\ }\emph {\bibinfo {title} {{Tailoring surface codes: Improvements
  in quantum error correction with biased noise}}},\ \href {\doibase
  10.25910/x8xw-9077} {Ph.D. thesis},\ \bibinfo  {school} {University of
  Sydney} (\bibinfo {year} {2020}),\ \bibinfo {note} {(qecsim:
  \url{https://github.com/qecsim/qecsim})}\BibitemShut {NoStop}%
\bibitem [{\citenamefont {Bennett}\ \emph {et~al.}(1996)\citenamefont
  {Bennett}, \citenamefont {DiVincenzo}, \citenamefont {Smolin},\ and\
  \citenamefont {Wootters}}]{PhysRevA.54.3824}%
  \BibitemOpen
  \bibfield  {author} {\bibinfo {author} {\bibfnamefont {C.~H.}\ \bibnamefont
  {Bennett}}, \bibinfo {author} {\bibfnamefont {D.~P.}\ \bibnamefont
  {DiVincenzo}}, \bibinfo {author} {\bibfnamefont {J.~A.}\ \bibnamefont
  {Smolin}}, \ and\ \bibinfo {author} {\bibfnamefont {W.~K.}\ \bibnamefont
  {Wootters}},\ }\bibfield  {title} {\enquote {\bibinfo {title} {Mixed-state
  entanglement and quantum error correction},}\ }\href {\doibase
  10.1103/PhysRevA.54.3824} {\bibfield  {journal} {\bibinfo  {journal}
  {Physical Review A}\ }\textbf {\bibinfo {volume} {54}},\ \bibinfo {pages}
  {3824} (\bibinfo {year} {1996})}\BibitemShut {NoStop}%
\bibitem [{\citenamefont {Kolmogorov}(2009)}]{kolmogorov2009blossom}%
  \BibitemOpen
  \bibfield  {author} {\bibinfo {author} {\bibfnamefont {V.}~\bibnamefont
  {Kolmogorov}},\ }\bibfield  {title} {\enquote {\bibinfo {title} {{Blossom V:
  a new implementation of a minimum cost perfect matching algorithm}},}\ }\href
  {\doibase 10.1007/s12532-009-0002-8} {\bibfield  {journal} {\bibinfo
  {journal} {Mathematical Programming Computation}\ }\textbf {\bibinfo {volume}
  {1}},\ \bibinfo {pages} {43} (\bibinfo {year} {2009})}\BibitemShut {NoStop}%
\bibitem [{\citenamefont {Higgott}(2021)}]{higgott2021pymatching}%
  \BibitemOpen
  \bibfield  {author} {\bibinfo {author} {\bibfnamefont {O.}~\bibnamefont
  {Higgott}},\ }\bibfield  {title} {\enquote {\bibinfo {title} {Pymatching: A
  python package for decoding quantum codes with minimum-weight perfect
  matching},}\ }\href {https://doi.org/10.48550/arXiv.2105.13082} {\bibfield
  {journal} {\bibinfo  {journal} {arXiv preprint arXiv:2105.13082}\ } (\bibinfo
  {year} {2021})}\BibitemShut {NoStop}%
\bibitem [{\citenamefont {Fowler}\ \emph
  {et~al.}(2012{\natexlab{b}})\citenamefont {Fowler}, \citenamefont
  {Whiteside},\ and\ \citenamefont {Hollenberg}}]{PhysRevLett.108.180501}%
  \BibitemOpen
  \bibfield  {author} {\bibinfo {author} {\bibfnamefont {A.~G.}\ \bibnamefont
  {Fowler}}, \bibinfo {author} {\bibfnamefont {A.~C.}\ \bibnamefont
  {Whiteside}}, \ and\ \bibinfo {author} {\bibfnamefont {L.~C.~L.}\
  \bibnamefont {Hollenberg}},\ }\bibfield  {title} {\enquote {\bibinfo {title}
  {Towards practical classical processing for the surface code},}\ }\href
  {\doibase 10.1103/PhysRevLett.108.180501} {\bibfield  {journal} {\bibinfo
  {journal} {Phys. Rev. Lett.}\ }\textbf {\bibinfo {volume} {108}},\ \bibinfo
  {pages} {180501} (\bibinfo {year} {2012}{\natexlab{b}})}\BibitemShut
  {NoStop}%
\bibitem [{\citenamefont {Earl}\ and\ \citenamefont
  {Deem}(2005)}]{earl2005parallel}%
  \BibitemOpen
  \bibfield  {author} {\bibinfo {author} {\bibfnamefont {D.~J.}\ \bibnamefont
  {Earl}}\ and\ \bibinfo {author} {\bibfnamefont {M.~W.}\ \bibnamefont
  {Deem}},\ }\bibfield  {title} {\enquote {\bibinfo {title} {Parallel
  tempering: Theory, applications, and new perspectives},}\ }\href
  {https://pubs.rsc.org/en/content/articlelanding/2005/CP/b509983h} {\bibfield
  {journal} {\bibinfo  {journal} {Physical Chemistry Chemical Physics}\
  }\textbf {\bibinfo {volume} {7}},\ \bibinfo {pages} {3910} (\bibinfo {year}
  {2005})}\BibitemShut {NoStop}%
\bibitem [{git()}]{git}%
  \BibitemOpen
  \href {https://github.com/QEC-project-2020/EWD-QEC} {\bibinfo  {journal}
  {github.com/QEC-project-2020/EWD-QEC}\ }\BibitemShut {NoStop}%
\bibitem [{\citenamefont {Bravyi}\ and\ \citenamefont
  {Vargo}(2013)}]{PhysRevA.88.062308}%
  \BibitemOpen
\bibfield  {journal} {  }\bibfield  {author} {\bibinfo {author} {\bibfnamefont
  {S.}~\bibnamefont {Bravyi}}\ and\ \bibinfo {author} {\bibfnamefont
  {A.}~\bibnamefont {Vargo}},\ }\bibfield  {title} {\enquote {\bibinfo {title}
  {Simulation of rare events in quantum error correction},}\ }\href {\doibase
  10.1103/PhysRevA.88.062308} {\bibfield  {journal} {\bibinfo  {journal} {Phys.
  Rev. A}\ }\textbf {\bibinfo {volume} {88}},\ \bibinfo {pages} {062308}
  (\bibinfo {year} {2013})}\BibitemShut {NoStop}%
\bibitem [{\citenamefont {Srivastava}\ \emph {et~al.}(2021)\citenamefont
  {Srivastava}, \citenamefont {Kockum},\ and\ \citenamefont
  {Granath}}]{srivastava2021xyz2}%
  \BibitemOpen
  \bibfield  {author} {\bibinfo {author} {\bibfnamefont {B.}~\bibnamefont
  {Srivastava}}, \bibinfo {author} {\bibfnamefont {A.~F.}\ \bibnamefont
  {Kockum}}, \ and\ \bibinfo {author} {\bibfnamefont {M.}~\bibnamefont
  {Granath}},\ }\href@noop {} {\enquote {\bibinfo {title} {{The XYZ$^2$
  hexagonal stabilizer code}},}\ } (\bibinfo {year} {2021}),\ \Eprint
  {http://arxiv.org/abs/2112.06036} {arXiv:2112.06036} \BibitemShut {NoStop}%
\end{thebibliography}%

%\appendix

%\section{Markov chain Monte Carlo for biased noise}

%\section{Pure Z noise on the XZZX code}

\end{document}